\documentclass[10pt]{article}

\usepackage[superscript,sort]{cite}

\usepackage{ifthen,ifpdf}
\ifpdf 
	\usepackage[pdftex]{hyperref}	\hypersetup{colorlinks=true,linkcolor=black,citecolor=black,urlcolor=blue,backref=page,bookmarks=true,breaklinks=true,plainpages=false }
	\usepackage[pdftex]{graphicx}
	\usepackage[usenames,pdftex]{color}
\else
	\usepackage[colorlinks=true,breaklinks=true,plainpages=false]{hyperref}
	\usepackage{graphicx}
	\usepackage[usenames]{color}
\fi

\usepackage{amsthm,amscd,amsxtra,amsfonts,amsmath,amssymb,multirow}
\usepackage{wrapfig}
\usepackage[footnotesize]{caption}
\usepackage[tiny,compact]{titlesec}
\usepackage[textwidth=0.8in,textsize=footnotesize]{todonotes}
\usepackage{algorithm,algorithmic,extarrows}

\begin{document}

\title{Multidimensional persistence in biomolecular data
}

\author{
Kelin Xia$^1$ and
Guo-Wei Wei$^{1,2,3}$ \footnote{ Address correspondences  to Guo-Wei Wei. E-mail:wei@math.msu.edu}\\
$^1$Department of Mathematics \\
Michigan State University, MI 48824, USA\\
$^2$Department of Electrical and Computer Engineering \\
Michigan State University, MI 48824, USA \\
$^3$Department of Biochemistry and Molecular Biology\\
Michigan State University, MI 48824, USA \\
}

\date{\today}
\maketitle

\begin{abstract}

Persistent homology has emerged as a popular technique for the topological simplification of big data, including biomolecular data. Multidimensional persistence bears considerable promise to bridge the gap between geometry and topology. However, its practical and robust construction has been a challenge. We introduce two families of   multidimensional persistence, namely  pseudo-multidimensional persistence and multiscale multidimensional persistence. The former is generated via the repeated applications of persistent homology filtration to high dimensional data, such as results from molecular dynamics or partial differential equations. The latter is constructed via isotropic and anisotropic  scales that create new simiplicial complexes and associated topological spaces. The utility, robustness and efficiency of the proposed topological methods are  demonstrated via protein folding, protein flexibility analysis, the topological denoising of cryo-electron microscopy data, and the scale dependence of nano particles. Topological transition between partial folded and unfolded proteins has been observed in multidimensional persistence. The separation between noise topological signatures and molecular topological fingerprints is achieved by the Laplace-Beltrami flow. The multiscale multidimensional persistent homology reveals relative local features in Betti-0 invariants and the relatively global characteristics of Betti-1 and Betti-2 invariants.

\end{abstract}


\newpage

{\setcounter{tocdepth}{5} \tableofcontents}

\newpage
  
\section{Introduction}

The rapid progress in science and technology has led to the explosion in biomolecular data.  The past decade  has witnessed a rapid growth in gene sequencing. Vast sequence databases are readily available for  entire genomes of many bacteria, archaea and eukaryotes. The  human genome decoding that originally took 10 years to process  can be achieved in a few  days nowadays.  The Protein Data Bank (PDB) updates new structures on a daily basis and  has  accumulated  more than  one hundred thousand tertiary structures.  The availability of these structural data enables the comparative  study of evolutionary processes,   gene-sequence based protein homology modeling of protein structures and the decryption  of the structure-function relationship. The abundant protein  sequence  and  structural information  makes it possible to build up  unprecedentedly comprehensive and accurate theoretical models. One of  ultimate goals is to predict protein functions from known protein sequences and structures, which remains a fabulous challenge.

Fundamental laws of physics described in quantum mechanics (QM), molecular mechanism (MM), continuum mechanics, statistical mechanics, thermodynamics, etc. underpin most physical models of biomolecular systems. QM methods are indispensable for chemical reactions, enzymatic processes and protein degradations  \cite{Cui:2002, YZhang:2009a}. MM approaches are able to elucidate the conformational landscapes  of proteins \cite{McCammon:1977}. However, both QM and MM involve an excessively large number of degrees of freedom and their application to real-time large-scale protein dynamics becomes prohibitively expensive. For instance, current computer simulations of protein folding take many months to come up with a very poor copy of what Nature administers  perfectly within a tiny fraction of a second. One way to reduce the number of degrees of freedom is to employ time-independent approaches, such as normal mode analysis (NMA)  \cite{Go:1983,Tasumi:1982,Brooks:1983,Levitt:1985}, flexibility-rigidity index (FRI) \cite{KLXia:2013d,Opron:2014} and  elastic network model (ENM) \cite{Tirion:1996}, including Gaussian network model (GNM)   \cite{Flory:1976, Bahar:1997,Bahar:1998} and  anisotropic network model (ANM) \cite{Atilgan:2001}. Another way is to incorporate continuum descriptions in atomistic representation to construct multiscale  models for large biological systems \cite{Warshel:1998,Sharp:1990a,Tully-Smith:1970, Cui:2002, YZhang:2009a,WFTian:2014,Geng:2011}. Implicit solvent models are some of the most popular approaches for solvation analysis \cite{Holst:1993,Baker:2004,Dong:2008MCB,Boschitsch:2004,Bertonati:2007,Georgescu:2002,Feig:2004b,Rocchia:2001,Chen:2008d,Zhou:2008b}. Recently,  differential geometry based multiscale models have been proposed for biomolecular structure, solvation, and transport \cite{Wei:2009,Wei:2012,DuanChen:2012a,Wei:2013}. The other way is to combine several atomic particles  into one or a few pseudo atoms or beads in coarse-grained (CG) models \cite{Rzepiela:2010,Smith:2001,FDing:2003,Paci:2002}. This approach is efficient for  biomolecular processes occurring at slow time scales and  involving large length scales.

All of the aforementioned theoretical models share a common feature: they are geometry based approaches \cite{ZYu:2008, XFeng:2012a,QZheng:2012} and depend on geometric modeling methodologies \cite{Sazonov:2012}. Technically, these approaches utilize geometric information,  namely, atomic coordinates,  angles, distances, areas \cite{Bates:2008,Bates:2009,QZheng:2012}  and sometimes curvatures \cite{ZhanChen:2010a,ZhanChen:2010b,ZhanChen:2012} as well as physical information, such as charges and their locations or distributions,  for the mathematical modeling of biomolecular systems.  Indeed, there is an increased   importance in geometric modeling for biochemistry \cite{XFeng:2012a}, biophysics \cite{XFeng:2013b,KLXia:2014a} and bioengineering \cite{Boileau:2013,Mikhal:2013}. Nevertheless, geometry based models  are typically   computationally expensive and become intractable for biomolecular processes such as protein folding, signal transduction, transcription and translation. Such a failure is often associated with massive data acquisition, curation, storage, search, sharing, transfer, analysis and visualization.   The  challenge originated from geometric modeling call for game-changing strategies, revolutionary  theories and  innovative methodologies.

Topological  simplification offers an entirely different strategy for big data analysis.  Topology deals with the connectivity of different components in a space and is able to classify independent entities,  rings and higher dimensional holes within the space. Topology captures geometric properties that are independent of metrics or coordinates. Indeed,  for many biological problems, including the  opening or closing of ion channels, the association or disassociation of ligands, and the assembly or disassembly of proteins,   it is the qualitative topology, rather than the quantitative geometry that determines physical and biological functions. Therefore, there is a   topology-function relationship  in many biological processes \cite{KLXia:2014c} such that topology is of  major concern.

 In contrast to geometric tools which are frequently inundated with too much structural information to be  computationally practical, Topological approaches often incur too much reduction of the geometric information.   Indeed, a coffee mug is topologically equivalent to a doughnut. Therefore, topology  is rarely used for quantitative modeling.  Persistent homology is a new branch of topology that is able to bridge the gap between traditional geometry and topology and provide a potentially  revolutionary approach to complex biomolecular systems.   Unlike  computational homology which gives rise to truly metric free or coordinate free representations, persistent homology is able to embed additionally geometric information into  topological invariants  via a filtration process so that ``birth"  and ``death" of  isolated components, circles, rings, loops, voids or cavities at all geometric scales  can be measured  \cite{Edelsbrunner:2002,Zomorodian:2005,Zomorodian:2008}. As such, the filtration process create  a multiscale representation of important topological features. Mathematically these topological features are described by  simplicial complexes, i.e., topological spaces  constructed by  points, line segments, triangles, and their higher-dimensional counterparts. The basic concept of persistent homology was introduced by Frosini and Landi~\cite{Frosini:1999} and  Robins~\cite{Robins:1999}, independently. The first realization was due to  Edelsbrunner et al.~\cite{Edelsbrunner:2002}. The concept was genearlized by  Zomorodian and Carlsson \cite{Zomorodian:2005}.  Many  efficient computational algorithms  have been proposed  in the past decade \cite{Bubenik:2007, edelsbrunner:2010,Dey:2008,Dey:2013,Mischaikow:2013}. Many methods have been developed for the geometric representation and visualization of topological invariants computed from persistent homology.   Among them, the barcode representation  \cite{Ghrist:2008} utilizes various horizontal line segments or bars to describe the ``birth'' and ``death'' of homology generators   over the filtration process. Additionally,  persistent diagram representation directly displays  topological connectivity in the filtration process.   The availability of efficient persistent homology tools \cite{javaPlex, Perseus} has led to  applications in a  diverse fields, including image analysis \cite{Carlsson:2008,Pachauri:2011,Singh:2008,Bendich:2010}, image retrieval \cite{Frosini:2013}, chaotic dynamics  \cite{Mischaikow:1999,Kaczynski:2004}, complex network \cite{LeeH:2012,Horak:2009},  sensor network \cite{Silva:2005}, data analysis \cite{Carlsson:2009,Niyogi:2011,BeiWang:2011,Rieck:2012,XuLiu:2012}, computer vision \cite{Singh:2008}, shape recognition \cite{DiFabio:2011}, computational biology \cite{Kasson:2007,Gameiro:2013,Dabaghian:2012,KLXia:2014c}, and nano particles  \cite{KLXia:2015a,BaoWang:2014}.  

The most successful applications of persistent homology have been limited to topological characterization identification and analysis (CIA). Indeed, there is little persistent homology based physical or mathematical modeling and quantitative prediction in the literature.  Recently, we have introduced persistent homology as unique means for the quantitative modeling and prediction of  nano particles, proteins and other biomolecules \cite{KLXia:2014c, KLXia:2015a}.  Molecular topological fingerprint (MTF), a recently introduced concept  \cite{KLXia:2014c}, is utilized not only for the CIA, but also for revealing topology-function relationships in protein folding and protein flexibility.  Persistent homology is found to provide excellent prediction of  stability and curvature energies for hundreds of nano particles  \cite{KLXia:2015a,BaoWang:2014}. Most recently, we have proposed a systematical variational framework to construct objective-oriented persistent homology (OPH) \cite{BaoWang:2014}, which is able to proactively extract desirable topological   traits from complex data. An example realization of the OPH is achieved via  differential geometry and Laplace-Beltrami flow \cite{BaoWang:2014}.  Most recently, we have developed persistent homology based topological denoising method for noise removal in volumetric data from cryo-electron microscopy (cryo-EM)    \cite{KLXia:2014arX}.  We have shown that persistent homology provides a powerful tool for solving ill-posed inverse problems in cryo-EM structure determination \cite{KLXia:2014arX}.

However, one dimensional (1D) persistent homology has its inherent limitations. It is suitable   for relatively simple systems described by one or a few parameters. The emergence of complexity in self-organizing biological systems frequently requires more comprehensive topological descriptions. Therefore, multidimensional persistent homology, or multidimensional persistence, becomes valuable for  biological systems as well as many other complex systems. 
In principle, multidimensional persistence should be able to seamlessly bridge geometry and topology.  Although multidimensional persistence bears  great promise, its construction is non-trivial and  elusive to the scientific community \cite{Carlsson:2009theory}. A major obstacle is that, theoretically, it has been proved there is no complete discrete representation for multidimensional persistent module analogous to one dimensional situation \cite{Carlsson:2009theory}. State differently, the persistent barcodes or persistent diagram representation is only available in one dimension filtration, no counterparts can be found in higher dimensions. Therefore, in higher dimensional filtration, incomplete discrete invariants that are computable, compact while still maintain important persistent information, are being considered \cite{Carlsson:2009theory}. Among them, a well-recognized one is persistent Betti numbers (PBNs) \cite{Edelsbrunner:2002}, which simply displays the histogram of Betti numbers over the filtration parameter.  The PBN is also known as rank invariant \cite{Carlsson:2009theory} and size functions (0th homology) \cite{Frosini:1999}.  A major merit of the PBN representation is its equivalent to the persistent barcodes in one dimension, which means that this special invariant is complete in 1D filtration.  Also, it has been proved that PBN is stable in the constraint of certain marching distance \cite{Cerri:2013}.  A few mathematical  algorithms have been  proposed \cite{Carlsson:2009computing,Cohen:2006,Cerri:2013}. Multi-filtration has been used in pattern recognition or shape comparison \cite{Frosini:1999,Biasotti:2008, Cerri:2013b}.  Computationally, the realization of  robust multidimensional persistent homology remains a challenge as algorithms proposed have to be topologically feasible, computationally efficient and practically useful.

The objective of this work is to introduce two classes of  multidimensional persistence for biomolecular data. One class of multidimensional persistence is generated by repeated applications of 1D persistent homology to high-dimensional data, such as those from protein folding, molecular dynamics, geometric partial differential equations (PDEs), varied signal to noise ratios (SNRs), etc. The resulting high-dimensional persistent homology is a  pseudo-multidimensional persistence. Another class of multidimensional persistence  is  created from a family of new simplicial complexes associated an isotropic scale or anisotropic scales. In general,   scales behave in the same manner as wavelet scales do. They can focus on the certain features of the interest and/or defocus on undesirable  characteristics. As a consequence, the proposed scale based isotropic and anisotropic filtrations give rise to new multiscale multidimensional persistence.   We demonstrate the application of the proposed   multidimensional persistence to a number of biomolecular and/or molecular systems, including protein flexibility analysis, protein folding characterization, topological denoising, noise removal from cryo-EM data, and analysis of fullerene molecules. Our multidimensional filtrations are carried out on  three types of data formats,  namely,  point cloud data, matrix data  and volumetric data. Therefore, the proposed methods can be easily applied to problems in other disciplines that involve similar data formats.

Our algorithm for  multidimensional persistence  is robust and straightforward. In a two-dimensional (2D)  filtration, we fix one of the filtration parameters and perform  the filtration on the second parameter to obtain PBNs. Then we systematically change the fixed parameter to sweep over its whole range, and stack all the PBNs together. This idea can be directly applied to three dimensional (3D) and higher dimensional filtrations. Essentially, we just repeat the 1D filtration over and over  until the full ranges of other parameters are sampled. The PBNs are then glued together. This multidimensional persistent homology method can be applied to any other high dimensional data. In this work,  point cloud data and matrix data are analyzed by using the JavaPlex \cite{javaPlex}. Volumetric data are processed with the Perseus  \cite{Perseus}.

The rest of this paper is organized as follows. In Section   \ref{sec:point_cloud_data}, we explore the multidimensional persistence in point cloud data for protein folding.  We model the protein unfolding process by an all-atom steer molecular dynamics (SMD). We consider both an all-atom representation and a coarse-grained representation to analyze  the SMD data. From our multifiltration analysis, it  is  found that PBNs associated with local hexagonal and pentagonal ring structures in protein  residues are preserved during the unfolding process while those due to global rings and cavities diminish. Coarse-grained representation is able to directly capture the dramatic  topological transition during the unfolding process. In Section \ref{sec:matrix_data}, we investigate the multidimensional persistence in matrix data. The  GNM Kirchhoff (or connectivity) matrix   and FRI correlation matrix are analyzed by multidimensional persistent homology. The present approach is able to predict the optimal cutoff distance of the GNM and the optimal scale of the FRI algorithm for  protein flexibility analysis. Section \ref{sec:volumetric_data} is devoted to the  multidimensional persistence in volumetric data. We analyze the  multidimensional topological fingerprints of Gaussian noise and demonstrate the  multidimensional topological denoising of synthetic data and cryo-EM data in conjugation with the Laplace-Beltrami flow method. Finally, we construct  multiscale 2D and 3D persistent homology to analyze the  intrinsic topological patterns of  C$_{60}$ molecule.  This paper ends with a conclusion.

\section{Multidimensional persistence in the point cloud data of protein folding}\label{sec:point_cloud_data}

In this section, we reveal multidimensional persistence in point cloud data associated with  protein folding process.  It is commonly believed that after the translation from mRNA,  unfolded polypeptide or random coil folds into a unique 3D structure which defines the protein function \cite{Anfinsen:1973}. However, protein folding does not always lead to a unique 3D structure. Aggregated or misfolded proteins are often associated with sporadic neurodegenerative diseases, such as mad cow disease, Alzheimer's disease and Parkinson's disease. Currently, there is no efficient means to characterize disordered proteins or  disordered aggregation, which is crucial to the understanding of the molecular mechanism  of degenerative disease. In this section, we show that  multidimensional persistence provides an efficient tool to characterize and visualize the orderliness of protein folding.

\subsection{Protein folding/unfolding processes}\label{sec:SMD}

The SMD  is commonly used to generate elongated protein configurations from its nature state \cite{Paci:2000,Hui:1998,Srivastava:2013}. Our goal is to examine the associated changes in the protein topological invariants induced by SMD.  There are three approaches to achieve SMD:  high temperature, constant force pulling, and constant velocity pulling \cite{Paci:2000,Hui:1998,Srivastava:2013}. Both implicit and explicit molecular dynamics can be used for SMD simulations.  The mechanical properties of protein FN-III$_{10}$ has been utilized to carefully design and valid SMD. Appropriate treatment of solvent environment in the implicit SMD is crucial. Typically,   a large box which can hold the stretched protein is required,  although the computational  cost is relatively high \cite{Gao:2002}. In our study, a popular SMD simulation tool \href{http://www.ks.uiuc.edu/Training/Tutorials/namd/namd-tutorial-html/}{NAMD} is employed to generate the partially folded and unfolded protein conformations. The procedure consists of  two steps: the relaxation of the given structure and unfolding simulation with constant velocity pulling. In the first step, the protein structure is downloaded from the Protein Data Bank (PDB), which  is the major reservoir for protein structures with atomic details. Then, the structure is prepared through the standard procedure, including adding missed hydrogen atoms, before  it is solvated with a water box which has an extra 5{\AA}~ layer, comparing with the initial minimal box that barely hold the protein structure \cite{KLXia:2014b}. The standard minimization and equilibration processes are carried out. We employ a total of 5000 time steps of equilibration iterations with the periodic boundary condition after 10000 time steps of initial energy minimization. In our simulations, we use a time increment of  2 femtoseconds (fs).  We set SMDk=7. The results are recorded after each 50 time steps, i.e., one frame for each 0.1 picosecond (ps). We accumulate  a total of 1000 frames or protein configurations, which are employed for our  persistent homology filtration. 

\subsection{All-atom and coarse-grained representations}\label{sec:models}

Persistent homology analysis of proteins can be carried out either in an all-atom representation or in CG representations \cite{KLXia:2014c}.  For the all-atom representation, various types of atoms, including O,  C, N, S, P, etc., are all included and regarded as equally important in our computation. We deliberately ignore the Hydrogen atoms in our structure during the filtration analysis, as we found that they tend to contaminate our local protein fingerprints.  The all-atom representation gives an atomic description of a given protein frame or configuration and is widely used in molecular dynamic simulation. In contrast,  CG representations describe  the protein molecule with the reduced number of degrees of freedom and are able to highlight important protein structure features. CG representations can be constructed in many ways. A standard coarse-grained representation of proteins used in our earlier topological analysis is to represent each  amino acid by the corresponding C$_{\alpha}$ atom \cite{KLXia:2014c}.  CG representations are efficient for describing large proteins and protein complexes and significantly reduce the cost of calculating topological invariants \cite{KLXia:2014c}.

\begin{figure}
\begin{center}
\begin{tabular}{c}
\includegraphics[width=0.8\textwidth]{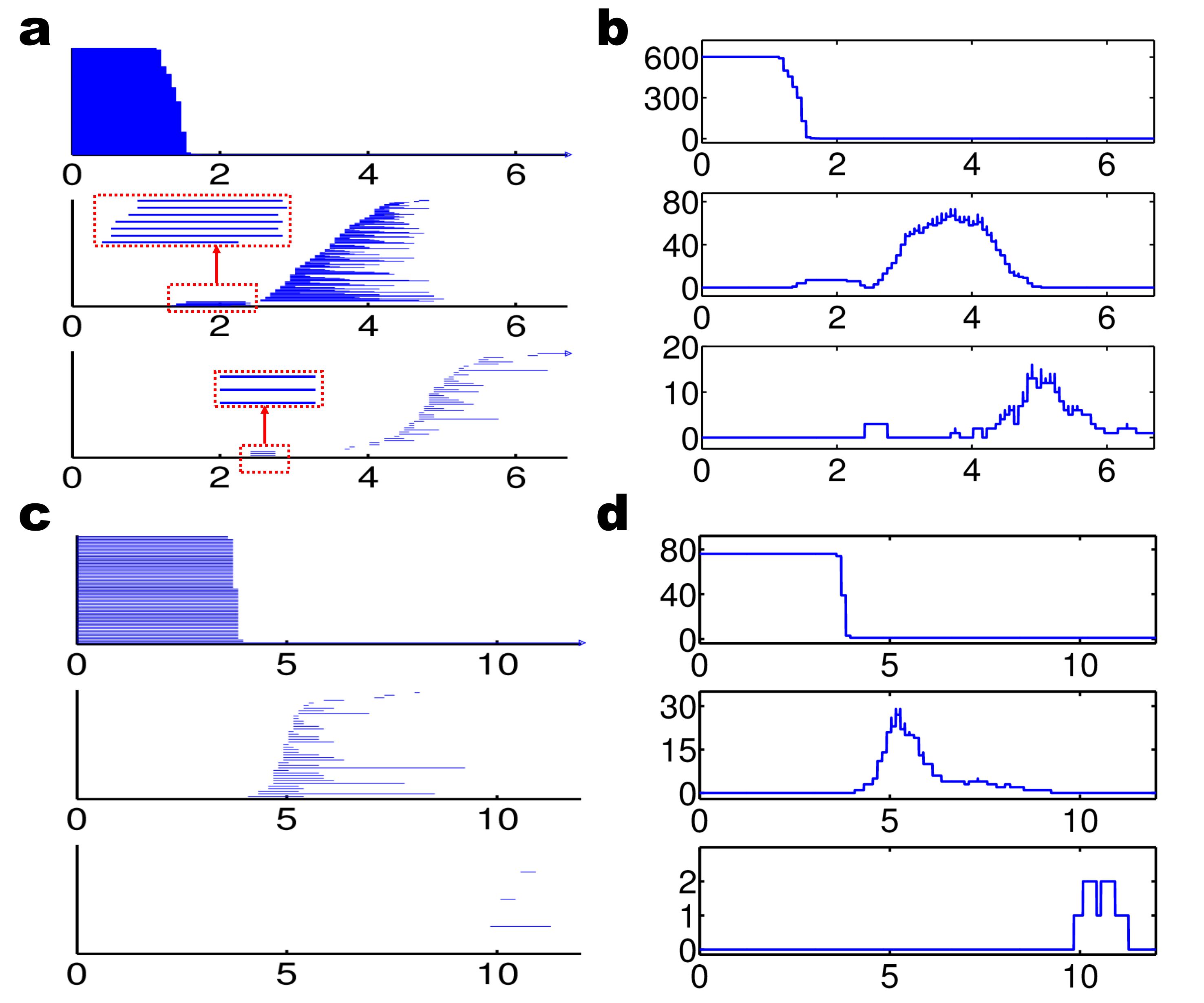}
\end{tabular}
\end{center}
\caption{ Persistent barcodes and PBNs  of 1UBQ   structure. 
{\bf a}  Persistent barcodes for the all-atom representation without hydrogen atoms;
{\bf b} PBNs for the all-atom representation without hydrogen atoms;
{\bf c}  Persistent barcodes for the CG representation;  
{\bf d}  PBNs for the CG representation. In each subfigure,  $\beta_0$,   $\beta_1$ and  $\beta_2$ are displayed in the top, middle and bottom panels, respectively. In all subfigures, horizontal axises label the filtration radius (\AA).  Vertical axises in {\bf b} and {\bf d} are the numbers of topological invariants. 
}
\label{fig:barcodes}
\end{figure}

Figure \ref{fig:barcodes} demonstrates the persistence information for the all-atom representation and the C$_{\alpha}$ coarse-grained representation of 1UBQ relaxation structure (i.e., the initial structure for the unfolding process). Figures \ref{fig:barcodes}{\bf a} and {\bf b} are  topological invariants from the all-atom representation without hydrogen atoms. In Fig. \ref{fig:barcodes}{\bf a}, it can be observed that $\beta_1$ and $\beta_2$ barcodes are clearly divided into two unconnected regions: local region (from 1.6 to 2.7 \AA~ for $\beta_1$ and from 2.4\AA~ to 2.7\AA~ for $\beta_2$) and global region (from 2.85\AA~ to 6.7\AA~ for $\beta_1$ and from 3.5\AA~ to 6.7\AA~ for $\beta_2$). Local region appears first during the filtration process and it is directly related to the hexagonal ring (HR) and pentagonal ring (PR) structures from the residues. As indicated in the zoomed-in regions enclosed by dotted red rectangles, there are 7 local $\beta_1$ bars and 3 $\beta_2$ bars, which are topological fingerprints for phenylalanine (one HR), tyrosine (one HR), tryptophan (one HR and one PR), proline (one PR), and histidine (one PR). In Fig. \ref{fig:folding_all_atom}{\bf a}, we have 3 hexagonal rings (red color) corresponding to 3 local $\beta_1$ bars  and 3 local $\beta_2$ bars.  It is well known that hexagonal structures produce $\beta_2$ invariants in the Vietoris$-–$Rips complex based filtration \cite{KLXia:2014c}. The other 4 local $\beta_1$ bars are from pentagonal structures (blue color). Figures \ref{fig:folding_all_atom} {\bf c} and {\bf d} are results from the coarse-grained representation. It can be seen that there is barely any   $\beta_2$ information for the initial structure. As protein unfolds, almost no cavities or holes are detected. Therefore, we only consider $\beta_0$ and $\beta_1$ invariants in the coarse-grained model.

\subsection{Multidimensional persistence in protein folding process}\label{sec:SMD1}

\begin{figure}
\begin{center}
\begin{tabular}{c}
\includegraphics[width=0.9\textwidth]{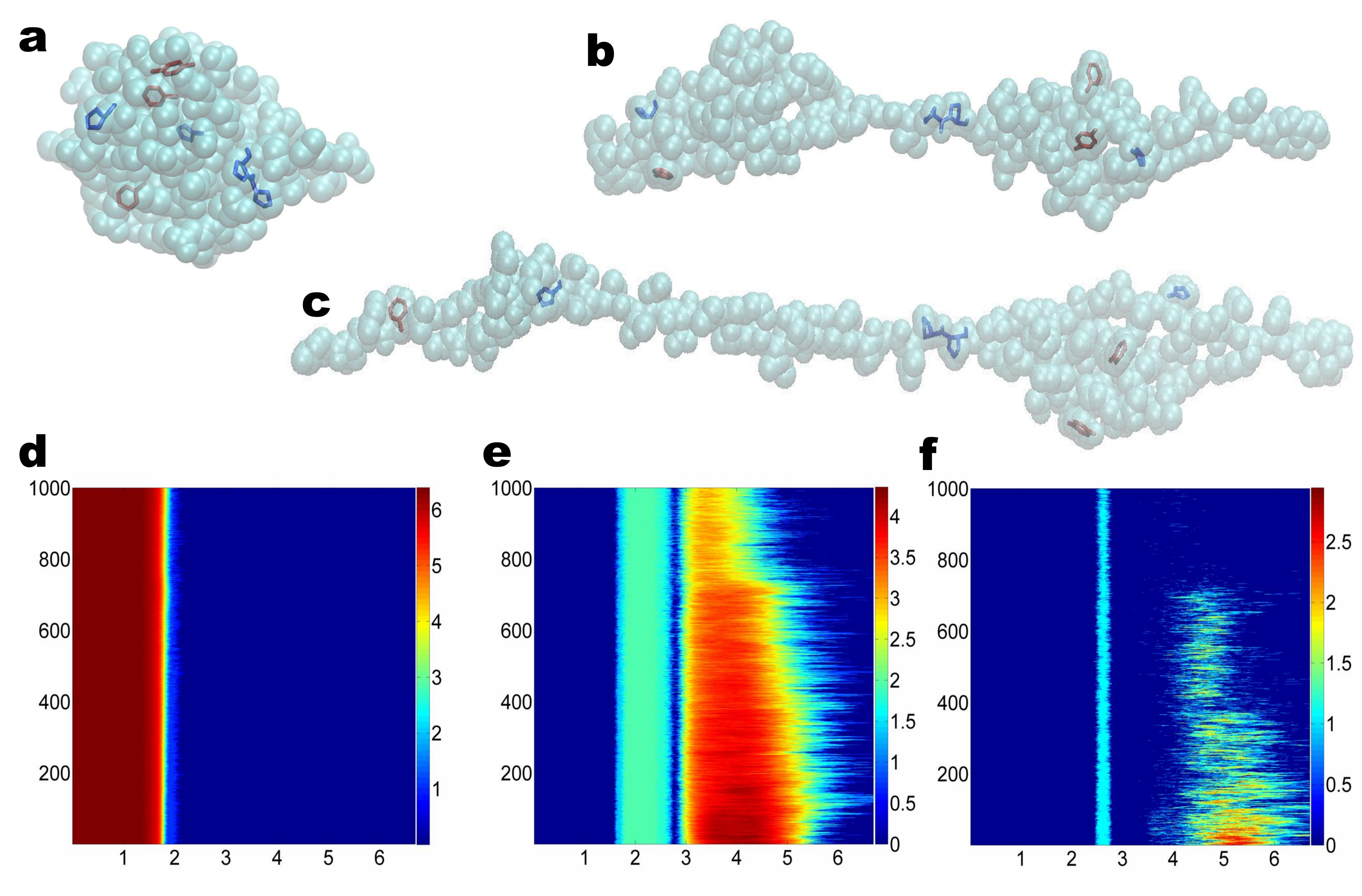}
\end{tabular}
\end{center}
\caption{The unfolding  of protein 1UBQ and the corresponding multidimensional persistence. 
{\bf a} All atom representation of the relaxed structure without hydrogen atoms; 
{\bf b} All atom representation of the unfolded structure at the 300th frame; 
{\bf c} All atom representation of the unfolded structure at the 500th frame; 
{\bf d} 2D $\beta_0$ persistence;  
{\bf e} 2D $\beta_1$ persistence;    
{\bf f} 2D $\beta_2$ persistence.   
In subfigures  {\bf d}, {\bf e} and {\bf f}, horizontal axises label the filtration radius (\AA) and the vertical axises are the  configuration index. The color bars denote the natural logarithms of PBNs.
}
\label{fig:folding_all_atom}
\end{figure}

In our  protein folding analysis, we extract 1000 configurations over the unfolding process. For each configuration, we carry out the point cloud filtration, i.e., systematically increasing the radius of ball associated with each atom,  and come up with three 1D PBN graphs for $\beta_0$, $\beta_1$, and $\beta_2$. We then stack 1000 PBN graphs of the same type, say all $\beta_0$ graphs,  together.  In this way, the final result can be stored in a 2D matrix with the row number indicating the filtration radius, the column number indicating the configuration, and the elements are PBN values.  Figures \ref{fig:folding_all_atom} {\bf a}, {\bf b} and {\bf c}  demonstrate the unfolding of protein  1UBQ in the  all atom representation without hydrogen atoms and the corresponding 2D persistence diagrams.  In  these subfigures,  we highlight residual pentagonal rings and hexagonal rings  in blue and red, respectively. These ring structures correspond to the local topological invariants as indicated in Fig. \ref{fig:barcodes}{\bf a}.   Figures \ref{fig:folding_all_atom} {\bf d}, {\bf e} and {\bf f} depict 2D persistent homology analysis of the protein unfolding process. Because all the bond lengths are around 1.5 to 2.0 \AA~ and do not change during the unfolding process, the 2D $\beta_0$ persistence shown in   Fig.  \ref{fig:folding_all_atom} {\bf d} is relatively simple and consistent with the top panel in Fig. \ref{fig:barcodes}{\bf b}. The 2D $\beta_1$ persistence shown in Fig.  \ref{fig:folding_all_atom} {\bf e} is very interesting. The local rings occurred from 1.6 to 2.7 \AA~ are due to  pentagonal   and hexagonal structures in the residues and are persistent over the unfolding process.   However, the numbers of $\beta_1$ invariants for global rings in the region  from 2.85 to 6.7 \AA~ vary dramatically during the unfolding process. Essentially, the SMD induced elongation of the polypeptide structure reduces the number of rings.  Finally, the behavior of the $\beta_2$ invariants in Fig.  \ref{fig:folding_all_atom} {\bf f}  is quite similar to that of the $\beta_1$. The local $\beta_2$ invariants occurred from  2.4 to 2.7 \AA~ induced by the hexagonal structures \cite{KLXia:2014c} remain unchanged during the unfolding process, while the number of global $\beta_2$ invariants occurred from 3.5 to 6.7 \AA~ rapidly decreases during the unfolding process. Especially, at 750th configuration and beyond, the number of $\beta_2$ invariants in global region has plummeted. The related PBNs for $\beta_2$ drop   to zero abruptly, indicates that the protein has become completely unfolded.   Indeed, there is an obvious topological transition in multidimensional $\beta_1$ persistence around 750th configuration as shown in    Fig.  \ref{fig:folding_all_atom} {\bf e}. The  global $\beta_1$ PBNs are dramatically reduced and their distribution regions are significantly narrowed  for all configurations beyond 75 picosecond simulations, which is an evidence for solely intraresidue $\beta_1$ rings.

\begin{figure}
\begin{center}
\begin{tabular}{c}
\includegraphics[width=0.8\textwidth]{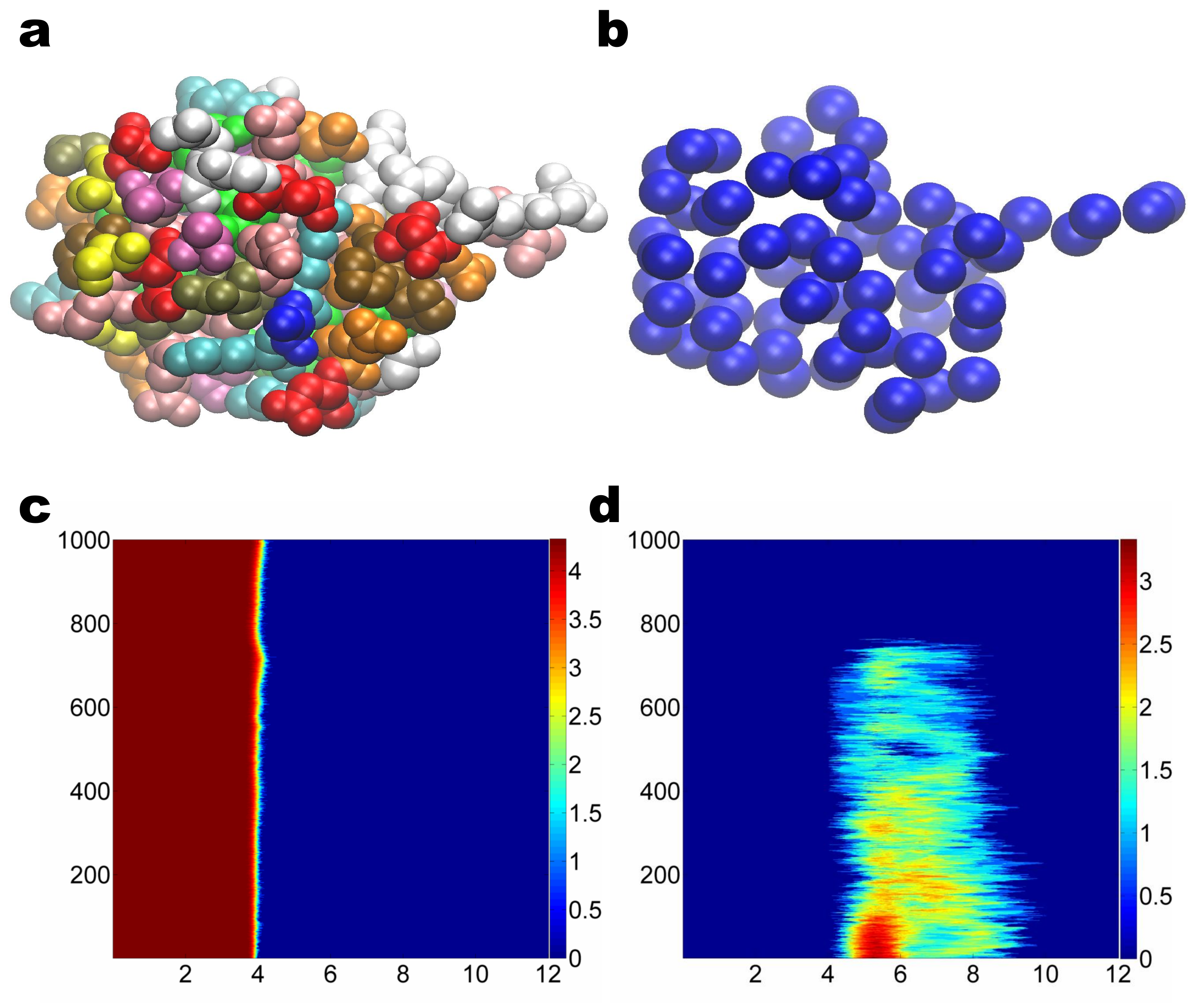}
\end{tabular}
\end{center}
\caption{Coarse-grained representation of the unfolding  of protein 1UBQ and the corresponding multidimensional persistence. 
{\bf a} All atom representation of the relaxed structure without hydrogen atoms; 
{\bf b} Coarse-grain representation of the relaxed structure without hydrogen atoms; 
{\bf c} 2D $\beta_0$ persistence;   
{\bf d} 2D $\beta_1$ persistence. 
The color in subfigure {\bf a} denotes different residues. 
In subfigures  {\bf c}, and {\bf d}, horizontal axises label the filtration radius (\AA) and the vertical axises are the configuration index. The color bars denote the natural logarithms of PBNs.
}
\label{fig:folding_ca}
\end{figure}

Having analyzed the multidimensional persistence in protein folding via the all-atom representation, it is interesting to further explore the same process and data set in the coarse-grained representation. Figure \ref{fig:folding_ca} illustrates our results. In  Fig. \ref{fig:folding_ca} {\bf a}, protein 1UBQ is plotted with all atoms except for hydrogen atoms. We use different colors to label different types of residues. The same structure is illustrated by the  C$_\alpha$ based CG representation in    Fig. \ref{fig:folding_ca} {\bf b}. An advantage of the CG model is that it simplifies topological relations by ignoring intraresidue topological invariants, while emphasizing interresidue topological features.    Figures \ref{fig:folding_ca} {\bf c}  and {\bf d} respectively depict 2D $\beta_0$ and $\beta_1$ invariants  of the protein unfolding process.  Compared with the all-atom results in Figs. \ref{fig:folding_all_atom} {\bf d} and {\bf e}, there are some unique properties. First, the CG analysis only emphasizes the global topological relations among residues and their evolution during the  protein unfolding. Additionally, the 2D $\beta_0$ profile of the all-atom representation was a strict invariant over the time evolution as shown in Fig. \ref{fig:folding_all_atom} {\bf d}, while that of CG model in Fig. \ref{fig:folding_ca} {\bf c}  varies obviously during the SMD simulation. The standard mean distance between two adjacent C$_{\alpha}$ atoms is about 3.8 \AA, which can be enlarged under the pulling force of the SMD. The deviation from the mean residue distance indicates the strength of the pulling force.    Finally, Fig. \ref{fig:folding_ca} {\bf d} displays a clear   topological transition from a partially folded state  to a completely  unfolded state at 75 picoseconds or 750th configuration. 
 
 As demonstrated  in our earlier work \cite{KLXia:2014c}, one can establish a quantitative model based on the PBNs of $\beta_1$ to predict the relative folding energy and stability.  The $\beta_1$ PBNs computed from the present CG representation are particularly suitable for this purpose. A similar quantitative model can be established to describe the orderliness of disordered proteins \cite{KLXia:2014c}. For simplicity, we omit these quantitative models and refer the interested reader to our earlier work  \cite{KLXia:2014c}.

In summary, multidimensional persistent homology analysis provides a wealth of information about protein folding and/or unfolding process including the number of atoms or residues, the numbers of  hexagonal rings and pentagonal rings in the protein, bond lengths or residue distances, the strength of applied pulling force, the orderliness of disordered proteins, the  relative folding energies, and topological translation  from partially folded states to completely unfolded states.  Therefore, multidimensional topological persistence is a powerful new tool for describing protein dynamics, protein folding and protein-protein interaction.

\section{Multidimensional persistence in  biological matrices}\label{sec:matrix_data}

Having illustrated the construction of multidimensional topological persistence in point cloud data, we further demonstrate the development of multidimensional topological analysis of matrix data. To this end, we consider biomolecular matrices associated flexibility analysis. The proposed method can be similarly applied to other biological matrices.

\subsection{Protein flexibility prediction}

Geometry,  electrostatics, and flexibility are some of the most important  properties for a protein that determine its functions.  The role of protein geometry and electrostatics has been extensively studied in the literature. However, the importance of protein flexibility is often overlooked. An interesting argument is that it is the protein flexibility, not disorder, that is intrinsic to molecular recognition \cite{Janin:2013}. Protein flexibility can be defined as its  ability to deform from the equilibrium state under external force.  The external stimuli are omnipresent either in the cellular environment and in the lattice condition. In response,  protein spontaneous fluctuations orchestrate with the Brownian dynamics in living cells or lattice dynamics in solid with its  degree of fluctuations determined by both the strength of external stimuli and protein flexibility. It has been shown that the Gaussian network model (GNM) and the flexibility-rigidity index (FRI)  are some of most successful methods  for protein flexibility analysis  \cite{KLXia:2013d,Opron:2014}. However, the performance of these methods depends on their parameters, namely, the cutoff distance of the GNM and the characteristic distance or the scale of the FRI.   In this work, we develop matrix based  multidimensional   persistent  homology methods to examine the optimal scale of FRI and optimal cutoff distance of the GNM. Brief descriptions are given to both methods to facilitate our persistent homology analysis.

\paragraph{Flexibility rigidity index}
The FRI have been proposed as a matrix diagonalization free method for the flexibility analysis of  biomolecules  \cite{KLXia:2013d,Opron:2014}.  The computational complexity of the fast FRI constructed by using the cell lists algorithm  is of $O(N)$, with $N$ being the number of particles \cite{Opron:2014}. In FRI, protein topological connectivity is measured by a correlation matrix.   Consider a protein with $N$ particles with coordinates given by   $\{ {\bf r}_{j}| {\bf r}_{j}\in \mathbb{R}^{3}, j=1,2,\cdots, N\}$. We denote $  \|{\bf r}_i-{\bf r}_j\|$ the Euclidean  distance between  $i$th  particle  and the $j$th  particle. For the $i$th particle, its correlation matrix element with the $j$th particles is given by  $ \Phi( \|{\bf r}_i - {\bf r}_j \|;\sigma_{j})$, where $\sigma_{j}$ is the  scale depending on the particle type.  The correlation matrix element is a   real-valued monotonically decreasing function satisfying
\begin{eqnarray}\label{eq:couple_matrix1-1}
\Phi( \|{\bf r}_i - {\bf r}_j \|;\sigma_{j})&=&1 \quad {\rm as }\quad  \|{\bf r}_i - {\bf r}_j \| \rightarrow 0\\ \label{eq:couple_matrix1-2}
\Phi( \|{\bf r}_i - {\bf r}_j \|;\sigma_{j})&=&0 \quad {\rm as }\quad  \|{\bf r}_i - {\bf r}_j \| \rightarrow\infty.
\end{eqnarray}
The  Delta sequences of the positive type discussed in an earlier work  \cite{GWei:2000} are suitable choices. For example,  one can select  generalized exponential  functions
\begin{eqnarray}\label{eq:couple_matrix1}
\Phi(\|{\bf r}_i - {\bf r}_j \|;\sigma_{j}) =    e^{-\left(\|{\bf r}_i - {\bf r}_j \|/\sigma_{j}\right)^\kappa},    \quad \kappa >0
\end{eqnarray}
and  generalized Lorentz functions
\begin{eqnarray}\label{eq:couple_matrix2}
 \Phi(\|{\bf r}_i - {\bf r}_j \|;\sigma_{j}) = \frac{1}{1+ \left( \|{\bf r}_i - {\bf r}_j \|/\sigma_{j}\right)^{\upsilon}},  \quad  \upsilon >0.
 \end{eqnarray}  
We have defined the atomic  rigidity index  $\mu_i$ for the $i$th particle  as \cite{KLXia:2013d}
\begin{eqnarray}\label{eq:rigidity1}
 \mu_i = \sum_{j}^N w_{j} \Phi( \|{\bf r}_i - {\bf r}_j \|;\sigma_{j} ), \quad \forall i =1,2,\cdots,N.
\end{eqnarray}
where $w_{j}$ is a particle  type dependent weight function. The  the atomic  rigidity index has a straightforward physical interpretation, i.e., a strong connectivity leads to a high  rigidity.  
 
We also defined the atomic flexibility index as the inverse of the atomic rigidity index,
\begin{eqnarray}\label{eq:flexibility1}
f_i= \frac{1}{\mu_i}, \quad \forall i =1,2,\cdots,N.
\end{eqnarray} 
The atomic flexibility indices $\{ f_i\}$ are used to predict experimental B-factors or  Debye-Waller factors via a linear regression \cite{KLXia:2013d}. The FRI theory has been intensively validated by a set of 365 proteins    \cite{KLXia:2013d,Opron:2014}.  It outperforms the GNM in terms of accuracy and efficiency  \cite{KLXia:2013d}.  
 
When we only consider one type of particles, say C$_\alpha$ atoms in a protein, we can set $w_{j}=1$. Additionally, it is convenient to set  $\sigma_{j}=\sigma$ for C$_\alpha$ based CG model. We use $\sigma$ as a scale parameter in  our multidimensional persistent homology analysis, which leads to a 2D persistent homology.

\paragraph{Elastic network model}

The  normal mode analysis (NMA)  \cite{Go:1983,Tasumi:1982,Brooks:1983,Levitt:1985} is a well developed technique and is constructed  based on the matrix diagonalization of MD  force field. It can be employed  to  study, understand and characterize the mechanical aspects of the long-time scale dynamics. The computational complexity for the matrix diagonalization is typically of $O(N^3)$, where $N$ is the number of matrix rows or particles.  Elastic network model (ENM) \cite{Tirion:1996} simplifies the MD force field by considering only the elastic interactions between nearby pairs of atoms. The Gaussian network model (GNM)   \cite{Flory:1976, Bahar:1997,Bahar:1998} makes a further simplification by using the coarse-grained  representation of a macromolecule.  This coarse-grained representation ensures  the computational efficiency.   Yang et al. \cite{LWYang:2008}   have  demonstrated  that the GNM is about one order more efficient than most other   matrix diagonalization based approaches. In fact, GNM is more accurate than the NMA \cite{Opron:2014}. 

The performance of GNM depends on its cutoff distance parameter, which allows only the nearby  neighbor atoms within the cutoff distance to be considered in the elastic Hamiltonian. In this work, we construct multidimensional persistent homology based on the cutoff distance in the GNM. We further analyze the parameter dependence of  the GNM by our 2D persistence.

\subsection{Persistent homology analysis of optimal cutoff distance}
\begin{figure}
\begin{center}
\begin{tabular}{c}
\includegraphics[width=0.8\textwidth]{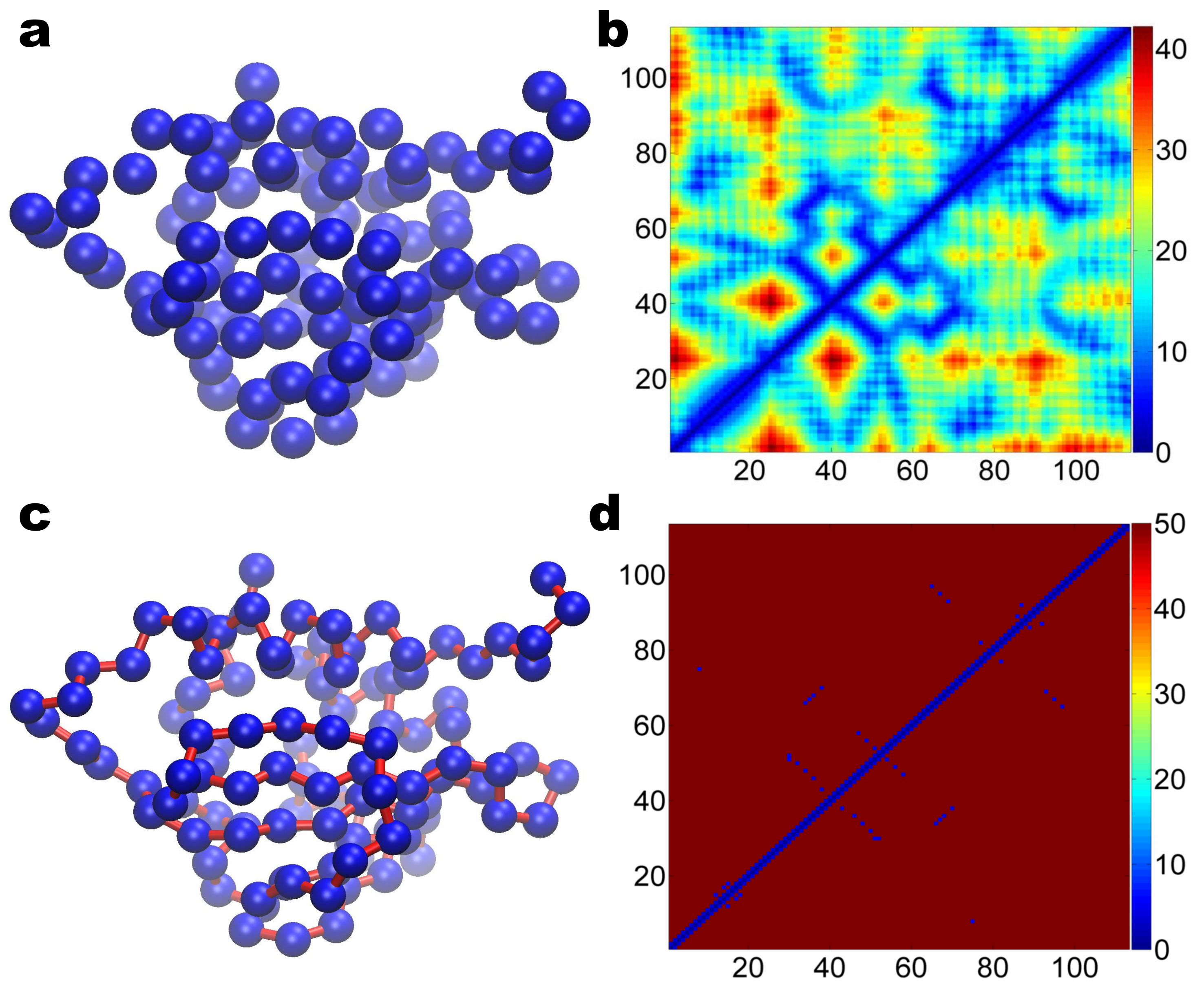}
\end{tabular}
\end{center}
\caption{Illustration of the elastic network, distance matrix and connectivity matrix of protein 1PZ4 in the coarse-grained representation. 
{\bf a} Coarse-grained C$_\alpha$ representation of protein 1PZ4  with 113 residues; 
{\bf b} Distance matrix of protein 1PZ4 at $r_c=50$\AA; 
{\bf c} Connectivity of protein 1PZ4 at $r_c=4$\AA;   
{\bf d} Distance matrix of protein 1PZ4 at $r_c=5$\AA.
In {\bf b} and {\bf d}  the horizontal and vertical axises are indices of residual numbers. 
The color bars in {\bf b} and {\bf d}  represent the distance between two residues. 
}
\label{fig:1pz4}
\end{figure}

Protein elastic network models, including the GNM,  usually employ the coarse-grained representation and do not distinguish between different residues.  Let us denote $N$ the total number of  C$_{\alpha}$ atoms in a protein, and $\|{\bf r}_i - {\bf r}_j \|$ the distance between $i$th  and $j$th C$_{\alpha}$ atoms.  To analyze the topological properties of protein elastic networks, we have introduced a new distance matrix ${\bf D} = \{D_{ij}|i=1,2,\cdots,N; j=1,2,\cdots,N\}$ \cite{KLXia:2014c} 
\begin{eqnarray}\label{eq:rigidity12}
 D_{ij} = \begin{cases}\begin{array}{lr}
     \|{\bf r}_i - {\bf r}_j \|, & \|{\bf r}_i - {\bf r}_j \|\leq r_c; \\
     d_\infty, & \|{\bf r}_i - {\bf r}_j \|>r_c,
\end{array}
\end{cases}
\end{eqnarray}
where $d_\infty$ is a sufficiently large value which is much larger than the final filtration size and  $r_c$ is a given cutoff distance. Here  $d_\infty$ is designed to ensure  that atoms beyond the cutoff distance $r_c$ do not form any high order simplicial complex during the filtration process.  Therefore, the resulting persistent homology shares the same topological connectivity  with  elastic network models.  Figure \ref{fig:1pz4} demonstrates the  construction of the  matrix representation for the Gaussian network model. Protein 1PZ4 with 113 residues is used as an example. Figure \ref{fig:1pz4} {\bf a} is  the coarse-grained C$_{\alpha}$ representation  of protein 1PZ4. Figure \ref{fig:1pz4} {\bf b} illustrates  the distance matrix ${\bf D}$ with a relatively large cutoff distance ($r_c=50$\AA). The $x$-axis and $y$-axis are residual number indices. The matrix value $D_{i,j}$ represents the distance between residue $i$ and residue $j$.  Figure \ref{fig:1pz4} {\bf c} is the connectivity of the Gaussian network model of protein 1PZ4 at $r_c=4$\AA. In this case, only the adjacent  C$_{\alpha}$ atoms are connected. The GNM distance matrix  at $r_c=5$\AA~ is illustrated in Fig. \ref{fig:1pz4} {\bf d}. By  systematically increasing the cutoff distance $r_c$, one can analyze the  topological connectivity and performance of the GNM. Additionally, the cutoff distance ($r_c$) in Eq. (\ref{eq:rigidity12}) is also employed as the filtration parameter in our 2D persistent homology analysis  of the GNM.

\begin{figure}
\begin{center}
\begin{tabular}{c}
\includegraphics[width=0.8\textwidth]{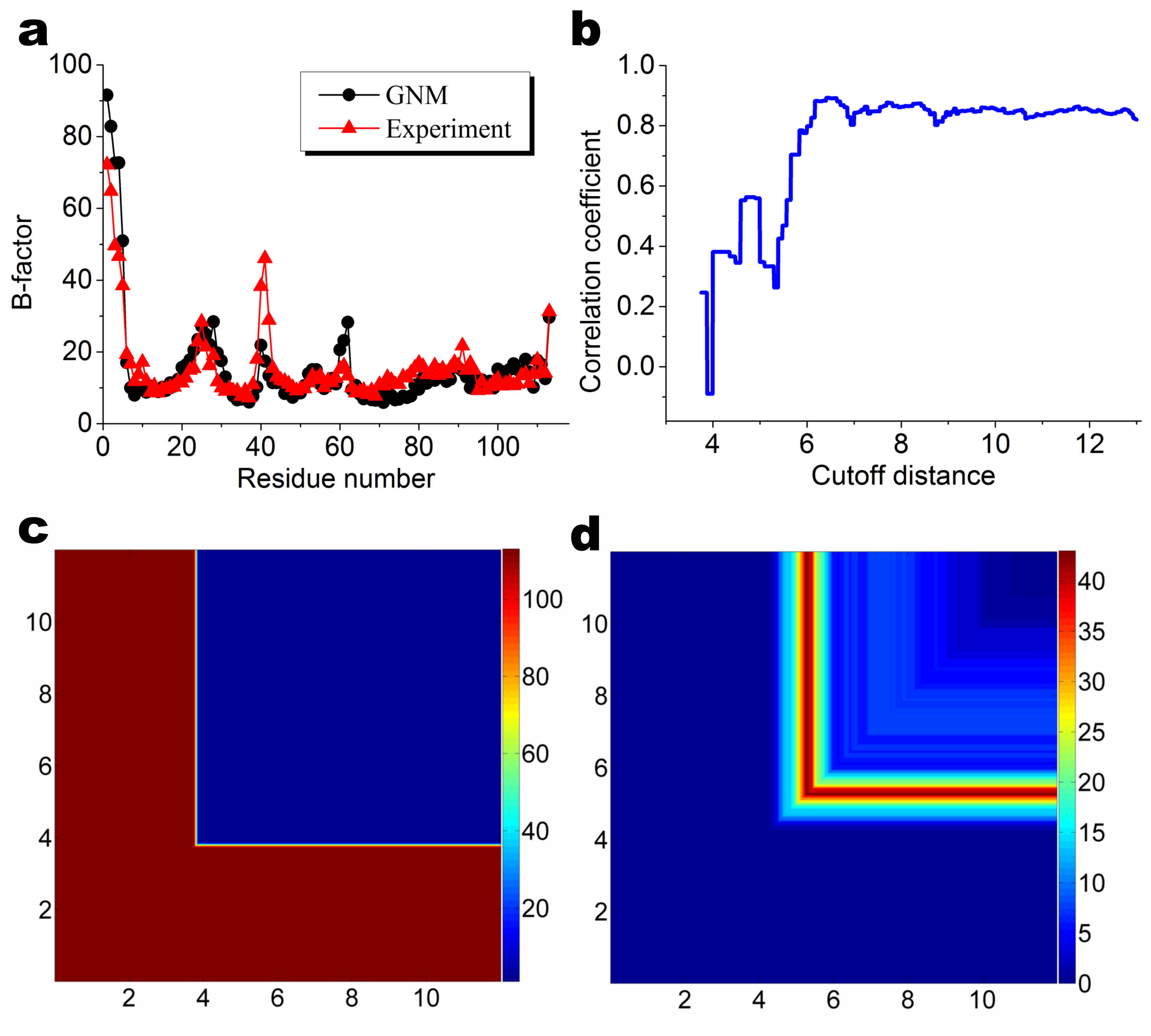}
\end{tabular}
\end{center}
\caption{Performance of GNM  and  multidimensional persistence of protein 1PZ4. 
{\bf a} Comparison of the GNM prediction at $r_c=6.6$\AA~ and experimental B-factors;
{\bf b} Correlation coefficient vs cutoff distance $(r_c)$ for the GNM; 
{\bf c} 2D $\beta_0$ persistence;  
{\bf d} 2D $\beta_1$ persistence.
In {\bf c} and {\bf d},   the horizontal axis is the cutoff distance $r_c$ in   filtration matrix (\ref{eq:rigidity12}) and the vertical axis  is the cutoff distance $r_c$ in the GNM. The color bars represent PBNs. }
\label{fig:1pz4top}
\end{figure}

The performance of the GNM for the B-factor prediction  and the multidimensional persistent homology analysis of protein 1PZ4 are plotted in  Fig. \ref{fig:1pz4top}. In Fig. \ref{fig:1pz4top} {\bf a}, we compare  the experimental B-factors and those predicted by the GNM  with a cutoff distance 6.6 \AA. The   Pearson correlation coefficient for the prediction is 0.89. The GNM provides   very good predictions except for the first three residues and the high flexibility around the 42nd residue.  Figure \ref{fig:1pz4top} {\bf b} shows the relation between correlation coefficient and cutoff distance. It can be seen that the largest correlation coefficients are obtained in the region when cutoff distance is in the range of 6\AA~ to  9\AA.  Figures \ref{fig:1pz4top} {\bf c} and {\bf d} illustrate 2D $\beta_0$ and $\beta_1$ persistence, respectively. The $x$-axises are the cutoff distance $r_c$ in  filtration matrix (\ref{eq:rigidity12}), which is the major filtration parameter.  The $y$-axises are the cutoff distance $r_c$ in  the GNM Kirchhoff  matrix. The resulting $\beta_0$ and $\beta_1$ PBNs in the matrix representation have unique patterns which  are highly symmetric along the diagonal lines. This symmetry, to a large extent, is duo to the way of forming  the GNM Kirchhoff matrix. The 2D $\beta_0$ persistence has an obvious interpretation in terms of 113 residues.  Interestingly, patterns in Fig. \ref{fig:1pz4top}   {\bf d} can be employed to explain the behavior of the correlation coefficients under different cutoff distances. To this end,  we roughly divide Fig. \ref{fig:1pz4top}  {\bf d} into four regions according to the cutoff distance, i.e.,  (0\AA, 4.5\AA), (4.5\AA, 5.8\AA), (5.8\AA, 9\AA)  and (9\AA, 12\AA). In the first region, the network is not well constructed. As the distance between two C$_{\alpha}$ atoms is around 3.8\AA, there is only a cluster of isolated atoms when cutoff distance is smaller than 4.5 \AA. Therefore, the corresponding GNMs do not give any  reasonable prediction.  In the second region, network structures begin to form. The number of 1D ring structures within these networks increases dramatically. It reaches its maximum when cutoff is about 5\AA, and then drops quickly. This behavior means that many local small-sized loops are developed. The corresponding GNMs can capture certain local properties, however, they neglect the global networks and are unable to grab the essential characteristics of the protein. As a consequence, the correlation coefficients are quite poor. In the third region,  constructed networks incorporate more and more large-sized loops or rings  and the corresponding GNMs improve predictions.   In the last region, local rings disappear while global rings are  included in  the network models. It is natural to assume that only when the constructed network  includes all essential topological invariants that the corresponding GNM  delivers the best prediction. However, this assumption turns out to be incorrect. As indicated in Fig. \ref{fig:1pz4top}  {\bf b}, the largest correlation coefficient is actually in the third region. The best cutoff distances are around 7\AA~ to 9\AA. This happens because  in the GNM, equal weights are assigned to all elastic springs once  spring lengths  are within the cutoff distance. Thus, there is  no discrimination between local and global ring structures.


\subsection{Persistent homology analysis of the FRI scale }

\begin{figure}
\begin{center}
\begin{tabular}{c}
\includegraphics[width=0.845\textwidth]{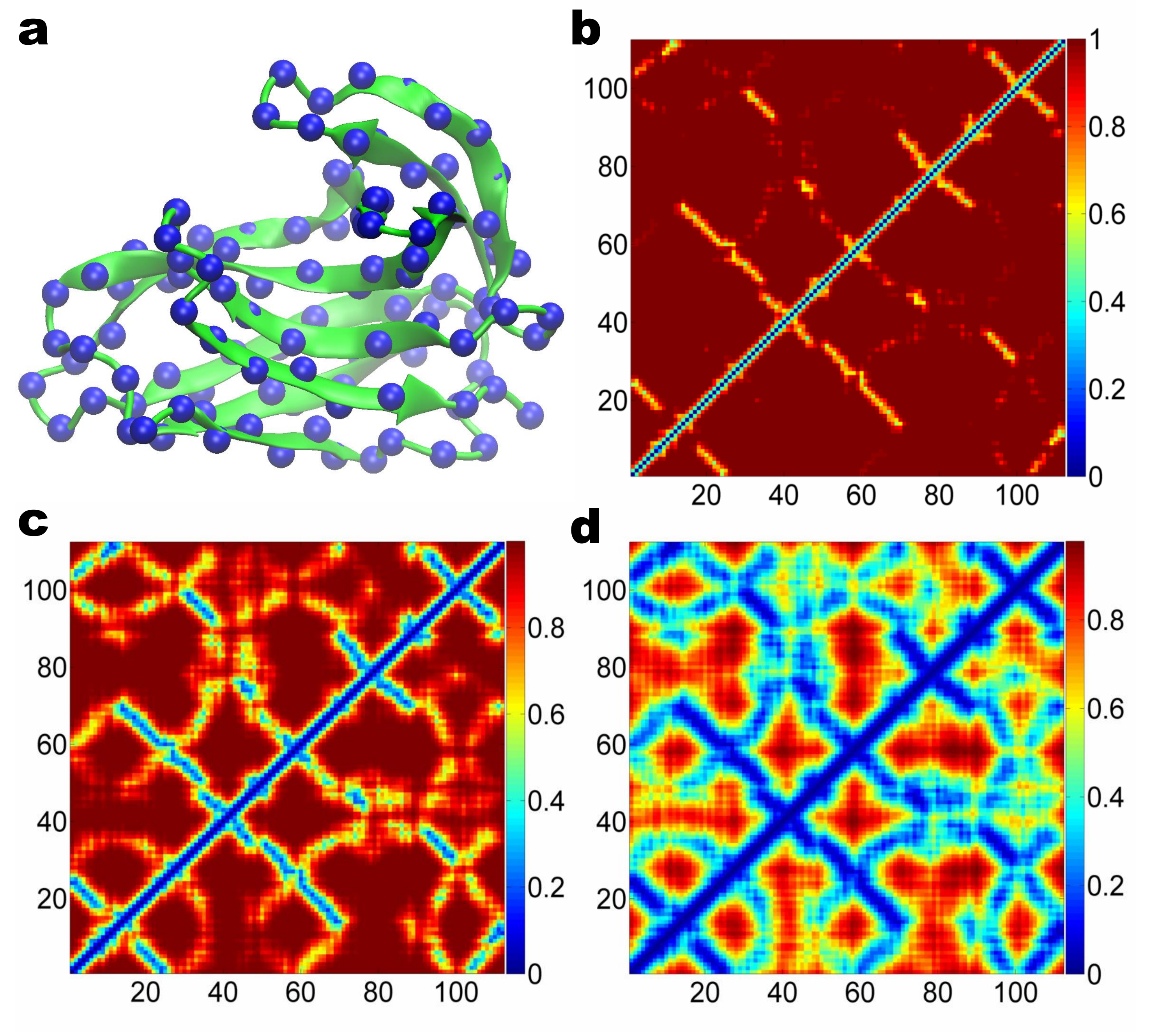}
\end{tabular}
\end{center}
\caption{ The coarse-grained representation and corresponding FRI filtration matrices of 2MCM. 
{\bf a} Coarse-grained model of 2MCM with 112 residues;
{\bf b} Filtration matrix at $\sigma=5$\AA; 
{\bf c} Filtration matrix at $\sigma=10$\AA;
{\bf d} Filtration matrix at $\sigma=20$\AA. 
In  {\bf b}, {\bf c} and {\bf d},  the horizontal and vertical axises represent residual number index. The color bars represent the values of the filtration matrices in Eq. (\ref{eq:couple_matrix01}). 
}
\label{fig:2mcm_matrix}
\end{figure}

Unlike GNM which utilizes  a  cutoff distance, the    FRI theory employs   a scale  or characteristic distance $\sigma$ in its  correlation kernel. The scale  has a similar function as the scale in wavelet theory, and thus  it emphasizes the contribution from the given scale.  The FRI scale has a direct impact in the accuracy of protein B-factor prediction. Similar to the optimal cutoff distance in the GNM,  the best FRI  scale varies from protein to protein, although an optimal value can be found based on a statistical average over hundreds of proteins \cite{KLXia:2013d,Opron:2014}. In the present work, we use the scale as an additional variable  to construct  multidimensional persistent homology. 

In our recent work, we have introduced a FRI based filtration method to convert the point cloud data into matrix data \cite{KLXia:2014c}. In this approach,   we construct a new filtration matrix  ${\bf M} = \{M_{ij}|i=1,2,\cdots,N; j=1,2,\cdots,N\}$  
\begin{eqnarray}\label{eq:couple_matrix01}
{M}_{ij} = \begin{cases}\begin{array}{lr}
      1-\Phi( \|{\bf r}_i - {\bf r}_j \|;\sigma), &  \quad i \neq j, \\
      0, & \quad i=j,
			\end{array}
\end{cases}
\end{eqnarray}
where $0 \leq \Phi( \|{\bf r}_i - {\bf r}_j \|;\sigma)\leq 1$ is defined in Eqs. (\ref{eq:couple_matrix1-1}) and  (\ref{eq:couple_matrix1-2}). 
To avoid any confusion, we simply use the exponential kernel with parameter $\kappa=2$ in the present work.

Figure \ref{fig:2mcm_matrix} {\bf a} presents a coarse-grained model for protein 2MCM, which has 112 residues. 
Figures \ref{fig:2mcm_matrix} 
{\bf b}, {\bf c} and {\bf d} are corresponding  filtration matrices derived from the FRI theory with $\sigma$ being set to 5.0\AA, 10.0\AA~ and 20.0\AA,~ respectively.  
The filtration matrices are constructed as ${M}_{ij}=1.0-e^{-\left( \frac{\|{\bf r}_i - {\bf r}_j \|}{\sigma}\right)^2}$.  Clearly, the scale $\sigma$ controls the degree of  connectivity of C$_\alpha$ atoms. At a small scale ($\sigma=5$\AA), only the nearest  neighbor  C$_\alpha$ atoms are correlated. At a relatively large scale ($\sigma=10$\AA), more  neighbor  C$_\alpha$ atoms are correlated. At a large scale  ($\sigma=20$\AA), many non-local correlations are emphasized.

\begin{figure}
\begin{center}
\begin{tabular}{c}
\includegraphics[width=0.8\textwidth]{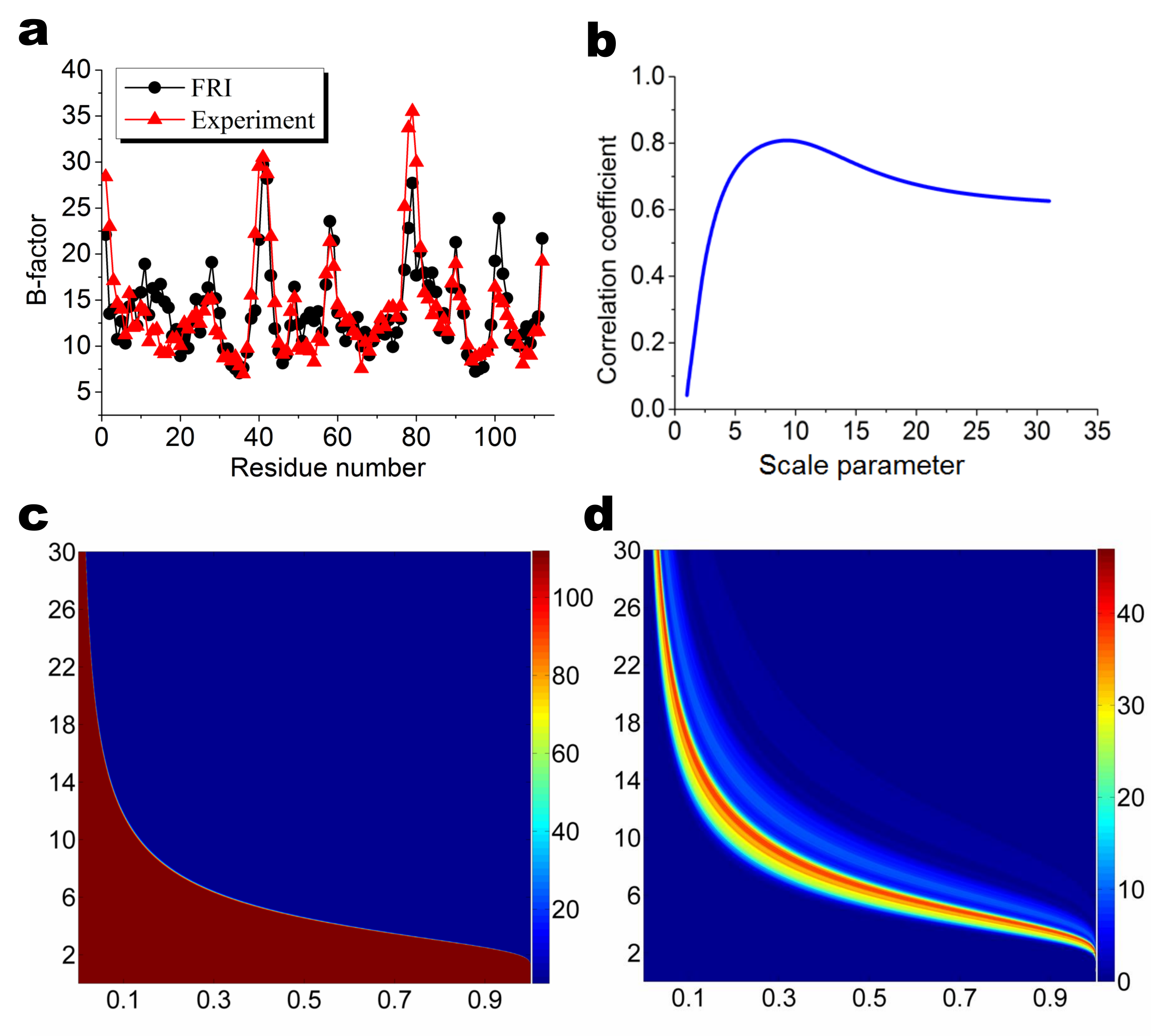}
\end{tabular}
\end{center}
\caption{Performance of the FRI and multidimensional persistence of protein 2MCM. 
{\bf a} Comparison of the FRI prediction at $\sigma=9.2$\AA~ and experimental B-factors;
{\bf b} Correlation coefficient vs scale ($\sigma$) for the FRI;
{\bf c} 2D $\beta_0$ persistence;  
{\bf d} 2D $\beta_1$ persistence. 
In {\bf c} and {\bf d}, the horizontal axis is in the FRI filtration matrix value $M_{ij}$ (\ref{eq:couple_matrix01}) and the vertical axis is the
 scale ($\sigma$) in terms of \AA~ in the FRI. The color bars represent PBNs.
}
\label{fig:2mcm}
\end{figure}

The performance of the FRI  B-factor prediction  and the multidimensional persistence of protein 2MCM are illustrated in Fig. \ref{fig:2mcm}. The  comparison of experimental B-factors and predicted B-factors with the scale $\sigma=9.2\AA$ is given in Fig. \ref{fig:2mcm} {\bf a}.  The Pearson  correlation coefficient is   0.81 for the prediction.   Figure \ref{fig:2mcm} {\bf b}, shows the relation between the correlation coefficient and the scale. It is seen that the largest correlation coefficients are obtained when the scale is in the range of   5\AA~ to 15\AA.  Figures \ref{fig:2mcm}  {\bf c} and {\bf d} demonstrate respectively   $\beta_0$ and $\beta_1$ 2D persistence.   Unlike the GNM results shown in Fig. \ref{fig:1pz4} where different cutoff distances lead to dramatic changes in network structures,  the FRI connectivity shown in  Fig. \ref{fig:2mcm}  {\bf c} increases gradually as $\sigma$ increases. For all $\sigma>3$\AA, the maximal $\beta_1$ values can reach 40 as shown in  Fig. \ref{fig:2mcm}  {\bf d}. However, in the region of 5\AA $<\sigma<$15\AA,  1D rings are established over a  wide range of the matrix values, which implies a wide range of distances.  The balance of the global and local rings gives rise to excellent FRI B-factor predictions.

In fact, a persistent homology based quantitative model can be established in terms of accumulated bar length \cite{KLXia:2014c}. Essentially, if all the PBNs  are added up at each scale, the accumulated PBNs give rise a good prediction of the optimal scale range. State differently, the plot of  the accumulated PBNs versus the scale will have a similar shape as the curve in      Fig. \ref{fig:2mcm} {\bf b}.

\section{Multidimensional  persistence in volumetric data}\label{sec:volumetric_data}

Volumetric data are widely available in science and engineering.  In biology, density information, such as the experimental data from   cryo-EM   \cite{Nogales:2006,KLXia:2014arX}, geometric flow based molecular hypersurface \cite{Bates:2008,Wei:2012,Wei:2013,BaoWang:2014} and electrostatic potential \cite{ZhanChen:2010a, DuanChen:2010b}, are typically described in volumetric form.   These  volumetric data can be filtrated directly in terms of isovalues  in persistent homology analysis.  Basically, the locations of the same density value form an isosurface. The discrete Morse theory can then be used to generate cell complexes. 
Additionally,  we have developed techniques  \cite{KLXia:2014c}  to convert point cloud data from X-ray crystallography into the volumetric form by using the rigidity function or density in our FRI algorithm \cite{KLXia:2014arX}.  Specifically,  the atomic  rigidity index  $\mu_i$  in Eq. (\ref{eq:rigidity1}) can be generalized to a position (${\bf r}$) dependent rigidity function or density   \cite{KLXia:2013d,Opron:2014}
\begin{eqnarray}\label{eq:rigidity3}
 \mu({\bf r}) & = & \sum_{j=1}^N w_{j}({\bf r}) \Phi( \|{\bf r} - {\bf  r}_j \|;\sigma_{j} ).
 \end{eqnarray}
Volumetric multidimensional persistence can be constructed in many different ways. Because $w_j$ and $\sigma_j$ are $2N$ independent variables, it is feasible to construct  $2N+1$-dimensional persistence for an $N$-atom biomolecule. Here the additional dimension is due to the filtration over the  density $\mu(\bf r)$.   If we set $w_j=1$ and $\sigma_j=\sigma$, we can construct genuine 2D persistence by filtration over two independent variables, i.e., $\sigma$ and density.

In this work, we also demonstrate the construction of   pseudo-multidimensional persistence.  Since noise and denoising are two important issues in  volumetric data, we develop methods for pseudo-multidimensional topological representation of noise and pseudo-multidimensional topological denoising.

\subsection{Multidimensional topological fingerprints of Gaussian noise}\label{Sec:HighOrderGF}

To analyze the topological signature of noise,  we make a case study on   Gaussian noise, which is perhaps the most commonly occurred noise. The Gaussian white noise is a set of   random events satisfying the normal distribution
\begin{eqnarray} \label{nosiedistrib}
n(t)=\frac{A_n}{ \sqrt{2\pi}\sigma_{n}}e^{-\frac{ (t-\mu_{n})^2}{2\sigma_{n}^2}},
\end{eqnarray}
where $A_n$, $\mu_{n}$ and $\sigma_{n}$ are the amplitude, mean value and standard deviation of the noise, respectively.  The strength of Gaussian white noise can be characterized by the signal to noise ratio (SNR) defined as  ${\rm SNR}=\mu_{s} / \sigma_{n}$, where  $\mu_s$ is the    mean value of signal. We generate noise polluted  volumetric data by adding different levels of Gaussian white noise to the original data.

\begin{figure}
\begin{center}
\begin{tabular}{c}
\includegraphics[width=0.9\textwidth]{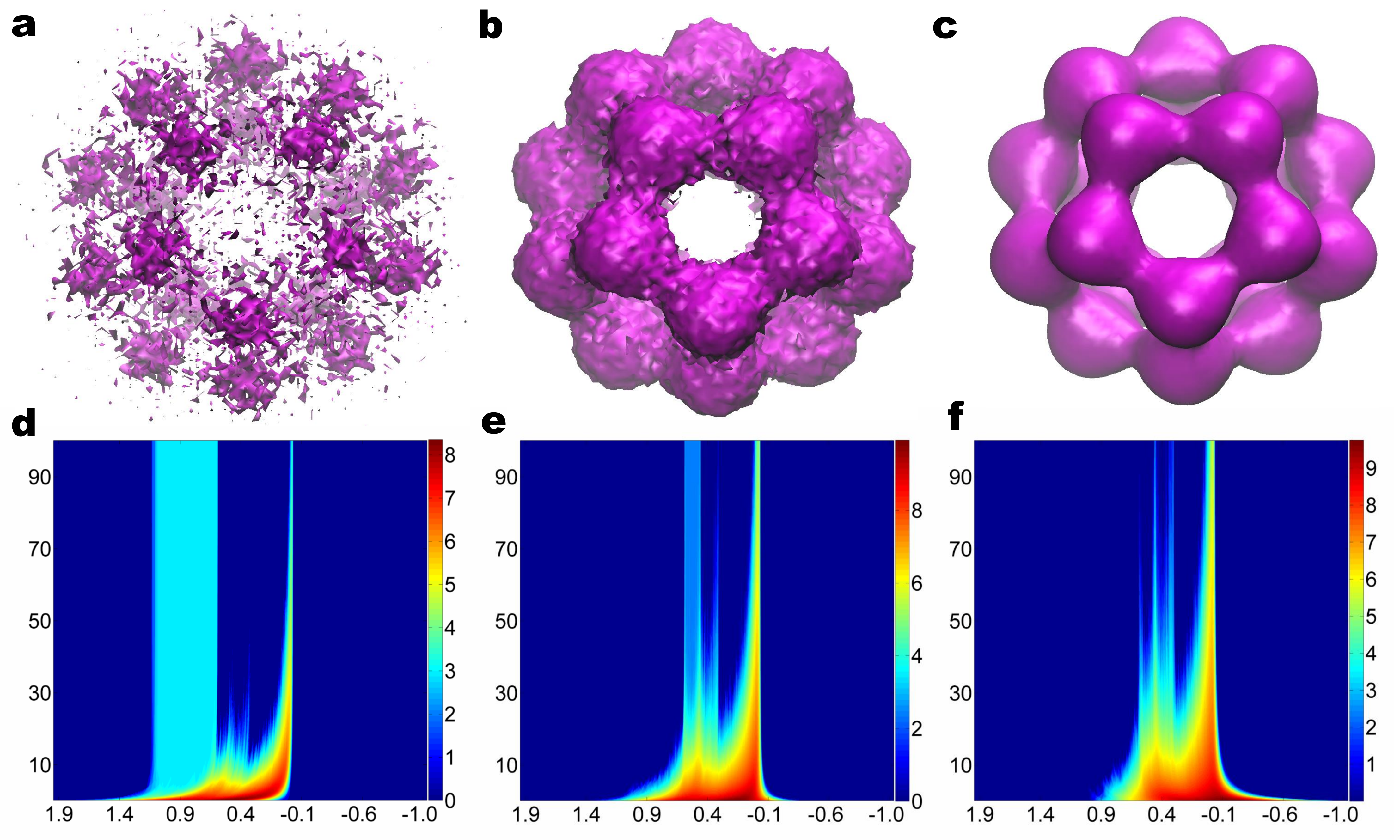}
\end{tabular}
\end{center}
\caption{Illustration of the  Gaussian noise contaminated fullerene  C$_{20}$ data and their multidimensional persistence. 
{\bf a} The   noisy  C$_{20}$ at the SNR of 1;
{\bf b} The   noisy  C$_{20}$ at the SNR of 10;
{\bf c} The   noisy  C$_{20}$ at the SNR of 100;
{\bf d} The 2D  $\beta_0$ persistence;
{\bf e} The 2D  $\beta_1$ persistence;  
{\bf f} The 2D  $\beta_2$ persistence. 
In {\bf d}, {\bf e}  and  {\bf f},  the horizontal axises represent the density isovalue (i.e., the main filtration parameter).  The vertical axises are the SNR. 
Color bars denote the natural logarithm of PBNs. 
}
\label{fig:noise}
\end{figure}

We employ fullerene C$_{20}$ as an example to illustrate the multidimensional topological fingerprints of noise. The rigidity density of     C$_{20}$ is given by 
\begin{eqnarray}\label{C20density}
\mu({\bf r}) & = & \sum_{j=1}^{20}  e^{- 2\|{\bf r} - {\bf r}_j \|}. 
\end{eqnarray}
The noisy data and multifiltration results are demonstrated in Fig. \ref{fig:noise}. We plot the noisy data of   C$_{20}$ with three SNRs, 1, 10 and  100 in Figs. \ref{fig:noise} {\bf a}, {\bf b} and {\bf c}, respectively. The persistent barcodes of C$_{20}$ have 20 $\beta_0$ bars, 11  $\beta_1$ bars and one  $\beta_2$ bar. Figures \ref{fig:noise} {\bf d}, {\bf e} and {\bf f} are respectively 2D $\beta_0$, $\beta_1$ and $\beta_2$ persistent homology. In these figures,  the vertical axises are the SNR values, which are varied over the range of  1.0 to 100.0. The horizontal axises represent the density isovalues (i.e., the main filtration parameter). In these cases, the designed filtration goes from the highest density value around 2.0 to the lowest density about -1.0.   The negative values are introduced by the Gaussian noise. The resulting PBNs are plotted in the  natural logarithm scale as indicated by the color bars.

First of all, the topological fingerprints of  C$_{20}$ stand out in Figs. \ref{fig:noise} {\bf d} and {\bf e} and demonstrate some invariant features as the SNR increases.   In Fig.  \ref{fig:noise} {\bf d}, the rectangle-like region   is due to the twenty isolated parts in C$_{20}$. Similarly,  the rectangle-like  region in Fig.  \ref{fig:noise} {\bf e} represents the 12 rings of the C$_{20}$ structure.  These rectangle patterns are the intrinsic topological fingerprints  of C$_{20}$. In Figs. \ref{fig:noise} {\bf d}, {\bf e} and {\bf f}, noise topological signatures  dominate the counts of Betti numbers, particularly when the SNR is smaller than 30. For example, $\beta_2$ spectrum near the density value of 0.4 is essentially indistinguishable from  noise induced cavities.

\subsection{Multidimensional topological denoising}\label{Sec:HighOrderGF2}

We have recently proposed topological denoising as a new strategy for topology-controlled noise reduction of synthetic, natural and experimental data \cite{KLXia:2014arX}. Our essential idea is to couple noise reduction with persistent homology analysis. Since persistent topology is extremely sensitive to the noise, the strength of noise signature can be monitored by persistent homology in a denoising process. As a result, one can make optimal decisions on number of deniosing iterations.  It was found that contrary to popular belief,  noise can have very long lifetimes in the barcode representation \cite{KLXia:2014arX},  while short lived features are part of molecular topological fingerprints \cite{KLXia:2014c}. In the present work, we introduce 2D topological denoising methods.  To this end, we present a brief review of the Laplace-Beltrami flow based denoising approach.

\paragraph{ Laplace-Beltrami flow}

One of efficient approaches for noise reduction in signals, images and data  is geometric analysis, which combines  differential geometry and differential equations. The resulting geometric PDEs have become very popular in applied  mathematics and computer science in the past two decades \cite{Mumford:1989, Willmore:1997, SOsher:1988}.   Wei introduced  some of the first families of high-order geometric PDEs for image analysis \cite{Wei:1999}
\begin{eqnarray}\label{eqn:highorder}
\frac{\partial u ({\bf r},t)}{\partial t}  =- \sum_{q}\nabla \cdot {\bf j}_q
+ e (u ({\bf r},t),|\nabla u ({\bf r},t)|, t), \quad q=0,1,2,\cdots
\end{eqnarray}
where the nonlinear hyperflux term ${\bf j}_q$ is given by
\begin{eqnarray}\label{eqn:highorder1-1}
{\bf j}_q=-
d_q (u ({\bf r},t),|\nabla u ({\bf r},t)|, t) \nabla \nabla^{2q}
u ({\bf r},t), \quad
q=0,1,2,\cdots
\end{eqnarray}
where ${\bf r}\in {\mathbb R}^n$, $ \nabla=\frac{\partial}{\partial{\bf r}} $, $u ({\bf r},t)$ is the processed signal, image or data,
$d_q (u ({\bf r},t),|\nabla u ({\bf r},t)|, t)$ are edge or gradient  sensitive diffusion coefficients
and $e (u ({\bf r},t),|\nabla u ({\bf r},t)|, t)$  is a nonlinear operator.
Denote  $X ({\bf r})$  the original noise data and set the initial input $u ({\bf r},0)=X ({\bf r})$.  There are many ways to choose hyperdiffusion coefficients  $d_q (u,|\nabla u|,t)$ in Eq.  (\ref{eqn:highorder1-1}). For example, one can use the exponential form
\begin{eqnarray}\label{eqn:highorder2}
d_q (u ({\bf r},t),|\nabla u ({\bf r},t)|,
t)=d_{q0}\exp\left[-\frac{|\nabla u|^\kappa}{\sigma_q^\kappa} \right],\quad \kappa>0,
\end{eqnarray}
where $d_{q0}$ is chosen as a constant with value depended on the noise level, and $\sigma_0$ and $\sigma_1$ are local statistical variance of $u$ and $\nabla u$
\begin{eqnarray}\label{eqn:highorder3}
\sigma_q^2 ({\bf r})= \overline{|\nabla^q u -\overline{\nabla^q
u}|^2} \quad  (q=0,1).
\end{eqnarray}
Here the notation $\overline{Y ({\bf r})}$ represents the local average of $Y ({\bf r})$ centered at position ${\bf r}$. The existence and uniqueness  of high-order geometric PDEs were investigated in the literature \cite{Greer:2004,Greer:2004b,MXu:2007,ZMJin:2010}. Recently, we have proposed differential geometry based objective oriented persistent homology to enhance or preserve desirable traits in the original data during the filtration process and then automatically detect or extract the corresponding topological features from the data \cite{BaoWang:2014}. From the point of view of signal processing,  the above high order geometric PDEs are designed as low-pass filters. Geometric PDE based high-pass filters was pioneered by Wei and Jia by  coupling two nonlinear geometric PDEs \cite{Wei:2002a}. Recently, this approach has been generalized  to a new formalism, the PDE transform, for signal,  image and data analysis \cite{YWang:2011c,YWang:2012b,YWang:2012d,QZheng:2012}.

Apart from their application to images   \cite{Wei:1999,MLysaker:2003,GGilboa:2004}, high order geometric PDEs have also been modified for macromolecular surface formation and evolution  \cite{Bates:2009},
 \begin{equation}\label{model4th3}
\frac{\partial S}{\partial t} = (-1)^q \sqrt{g ( |\nabla\nabla^{2q}
S|)} \nabla \cdot \left ( \frac{ \nabla  (\nabla^{2q} S)} { \sqrt{g (
|\nabla\nabla^{2q} S |)}} \right) + P (S,|\nabla S|),
\end{equation}
where $S$ is the hypersurface function,  $g (|\nabla\nabla^{2q}S|)=1+ |\nabla\nabla^{2q} S|^2$ is the generalized Gram determinant and $P$ is a generalized potential term. When $q=0$ and $P=0$, a Laplace-Beltrami equation is obtained \cite{Bates:2008},
 \begin{equation}\label{mean curvature_flow}
\frac{\partial S}{\partial t} =   |\nabla S| \nabla \cdot \left ( \frac{ \nabla S} {
|\nabla S |} \right).
\end{equation}
We employe this Laplace-Beltrami equation for  the noise removal in this work.

Computationally, the finite different method is used to discretize the Laplace-Beltrami equation in 3D. Suitable time interval $\delta t$ and grid spacing $h$ are required to ensure the stability and accuracy. To avoid confusion and control the noise reduction process systematically, we simply ignore the voxel spacing in different data sets and employ a set of unified parameters of $\delta t=5.0E-6$  and $h = 0.01$ in our computation. The intensity of noise reduction is then described by the duration of time integration or the number of iterations of Eq. (\ref{mean curvature_flow}).

\paragraph{Topological fingerprint identification}\label{Sec:Topological_fingerprint}

In the earlier example shown  in Fig. \ref{fig:noise}, it can been seen that, with the increase of SNR,  the intrinsic topological properties emerge and persist. Persistent patterns can be seen in PBNs, especially in $\beta_0$ and $\beta_1$.  It is interesting to know whether the topological persistence of the signal is a feature in the denoising process.

\begin{figure}
\begin{center}
\includegraphics[width=0.8\textwidth]{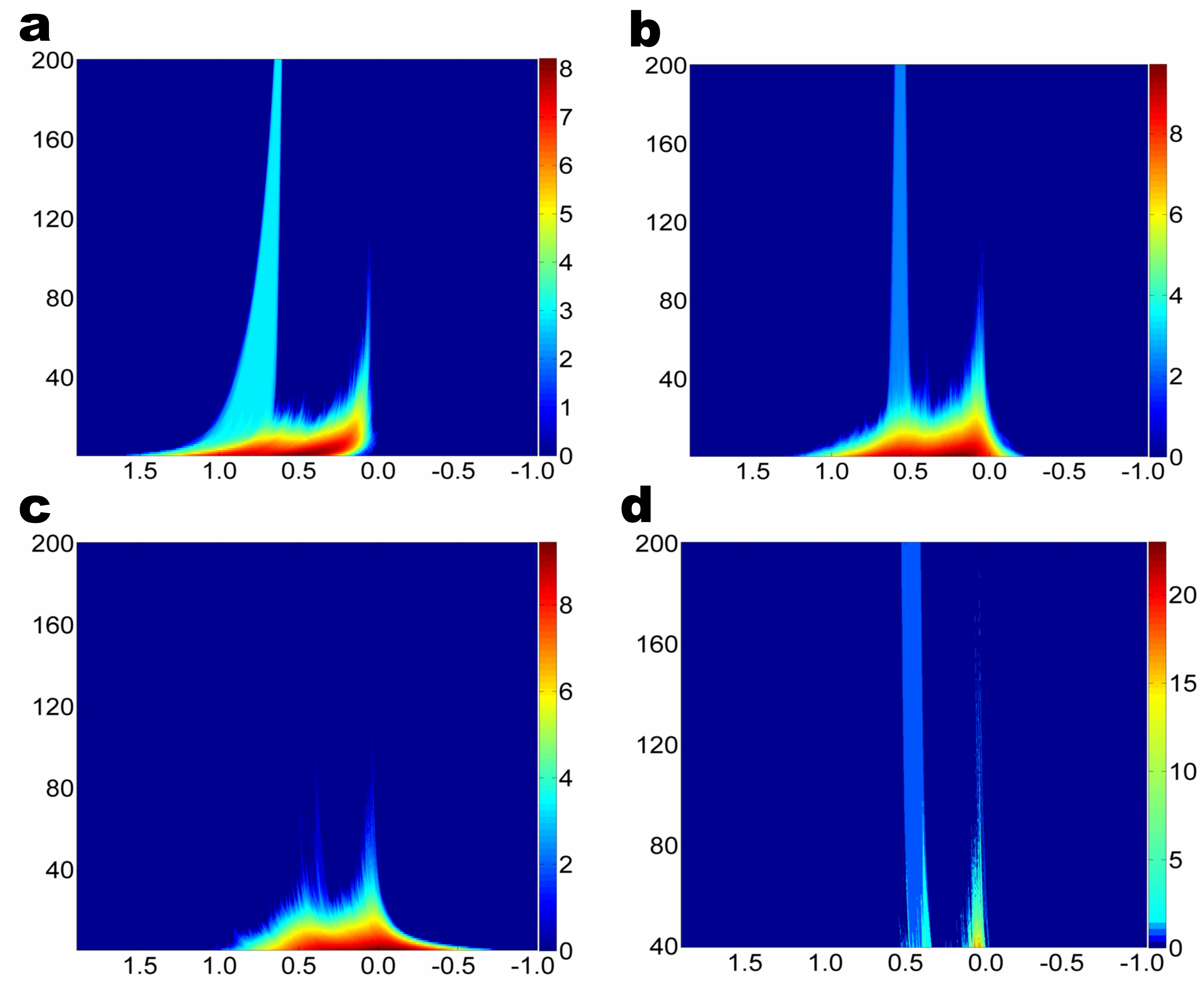}
\end{center}
\caption{ Multidimensional topological denoising for noise contaminated fullerene C$_{20}$ data with SNR 1.0. 
{\bf a} 2D $\beta_0$ persistence; 
{\bf b} 2D $\beta_1$ persistence; 
{\bf c} 2D $\beta_2$ persistence;  
{\bf d} A zoomed in picture of {\bf c}.  
In these plots,  the horizontal axises are for the density isovalue (i.e., the main filtration parameter).  The vertical axises represent the number of iterations. A total of 200 iterations is employed. Color bars denote the natural logarithm of PBNs in {\bf a}, {\bf b} and {\bf c}. Whereas, the color bar in {\bf d} represents the PBNs. 
}
\label{fig:denoise1}
\end{figure}

Figure \ref{fig:denoise1} depicts the topological invariants of contaminated fullerene C$_{20}$ over the  Laplace-Beltrami flow based denoising process.  The fullerene C$_{20}$ rigidity density is generated by using Eq.  (\ref{C20density}). The noise is added according to Eq. (\ref{nosiedistrib}) with the SNR of 1.0. The Laplace Beltrami  equation (\ref{mean curvature_flow}) is solved with time stepping  $\delta t=5.0E-6$  and spatial spacing $h = 0.01$.  Figures \ref{fig:denoise1}  {\bf a}, {\bf b}, and {\bf c} illustrate respectively the $\beta_0$,  $\beta_1$ and $\beta_2$ persistent homology analysis of the denoising process. The filtration goes from density 2.0 to $-1.0$.  A total of 200 denoising iterations are applied to the noisy data. The PBNs are plotted in the natural logarithm scale.   It can be seen that after about 40 denoising iterations, the noise intensity has been reduced dramatically. Indeed, the intrinsic topological features of  C$_{20}$ emerge and persist. It appears that the bandwidths of C$_{20}$ $\beta_0$ and $\beta_1$ topological fingerprints reduce  during the denoising process. However, such a bandwidth reduction is due to the fact that there is a dramatic   density reduction during the denoising precess, particularly at the early stage of the denoising. In fact,  the accumulated Betti numbers of C$_{20}$ do not change and stay stable. To support this claim, we highlight in {\bf d} the $\beta_2$ persistence of {\bf c} by ignoring the first 39 frames which are extremely noisy.  It should be noted  the color bar in {\bf c}, denotes PBNs in natural  logarithm scale, while the color bar  in {\bf d} is the PBN. As claimed, there is an excellent persistence of the $\beta_2$ invariant over the denoising process.

\begin{figure}
\begin{center}
\begin{tabular}{c}
\includegraphics[width=0.9\textwidth]{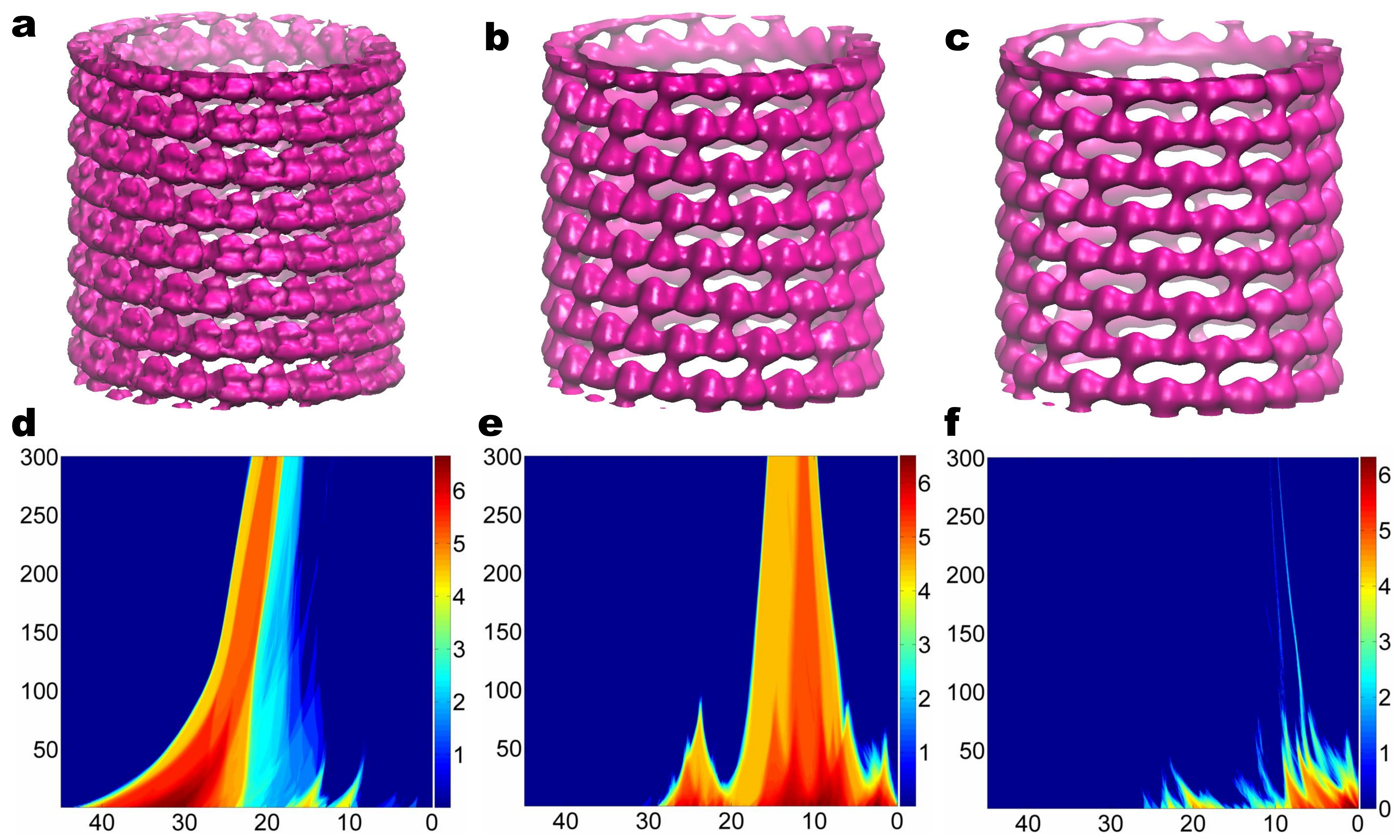}
\end{tabular}
\end{center}
\caption{Multidimensional topological denoising for  EMD 1129 data of a microtubule structure. 
{\bf a} Denoising data after one iteration; 
{\bf b} Denoising data after 100 iteration; 
{\bf c} Denoising data after 200 iteration;   
{\bf d} 2D $\beta_0$ persistence; 
{\bf e} 2D $\beta_1$ persistence;  
{\bf f} 2D $\beta_2$ persistence. 
Isosurfaces in {\bf a}, {\bf b} and  {\bf c} are extracted at isovalue  15.0.  In {\bf d}, {\bf e} and {\bf f},  the horizontal axises are   density isovalues (i.e., the main filtration parameter).  The vertical axises represent the number of iterations.   Color bars  denote the natural logarithm of PBNs. }
\label{fig:denoise2}
\end{figure}

Having demonstrated the construction of 2D persistence for topological denoising, we further apply this new technique for the analysis of noisy cryo-EM data of a microtubule (EMD 1129) \cite{Nogales:2006}. Figures \ref{fig:denoise2}  {\bf a}, {\bf b}, and {\bf c}  are surfaces  extracted from denoising data with the numbers of iterations of 1, 100 and 200, respectively.  A common isovalues of 15.0 used to extract surfaces in these plots.  It is seen that the denoising process reduces not only the noise, but also the density, which leads to the shift in the topological distribution.   Figures \ref{fig:denoise2} {\bf d}, {\bf e} and {\bf f} are respectively the 2D $\beta_0$, $\beta_1$ and $\beta_2$ persistence. The filtrations in horizontal axises go  from density 45 to 0. In Figs. \ref{fig:denoise2} {\bf d}, {\bf e} and {\bf f},  vertical axises are the numbers of iterations.  A total of 300 iterations is employed for integrating  Eq. (\ref{nosiedistrib}) with time stepping  $\delta t=2.0E-6$  and spatial spacing $h = 0.01$. Color bar values represent the natural logarithm of PBNs.  	It can be seen that after about 100 denoising iterations, the noise intensity has been  dramatically reduced. Persistent behavior can be observed in $\beta_0$, $\beta_1$ and $\beta_2$.  This persistent behavior is a manifest of  the intrinsic topological features of the micortubule structure.  

%

\subsection{Multiscale multidimensional  persistence }

In this section, we demonstrate the construction of multiscale multidimensional  persistent homology. To this end, we consider fullerene C$_{60}$, whose topological properties have been analyzed in our earlier work \cite{KLXia:2014c}, as an example to illustrate our method.   

\paragraph{Multiscale 2D persistence}

\begin{figure}
\begin{center}
\begin{tabular}{c}
\includegraphics[width=0.8\textwidth]{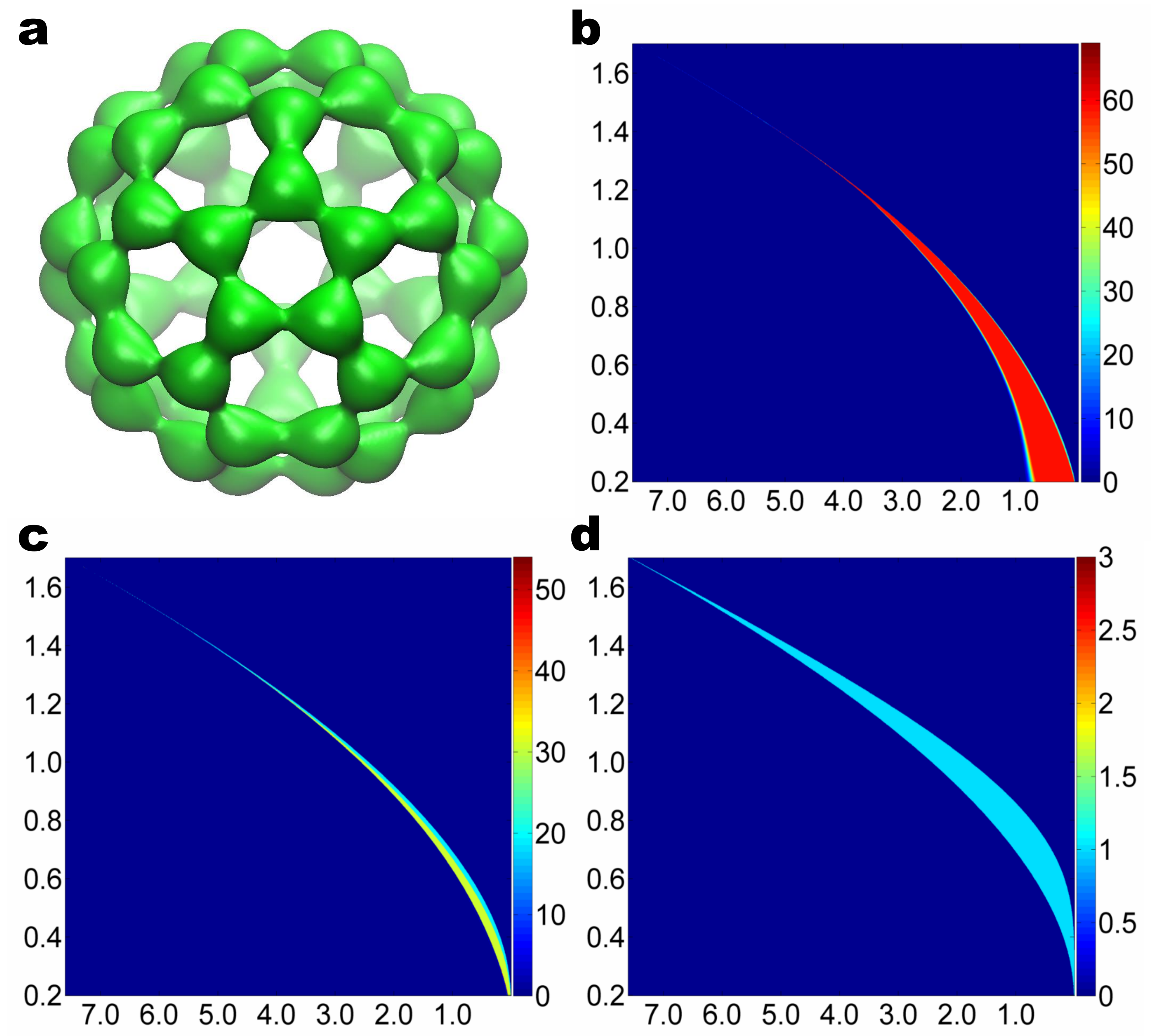}
\end{tabular}
\end{center}
\caption{The multiscale multidimensional persistence in fullerene C$_{60}$ volumetric data. 
{\bf a} Fullerene C$_{60}$ structure; 
{\bf b} 2D $\beta_0$ persistence; 
{\bf c} 2D $\beta_1$ persistence;  
{\bf d} 2D $\beta_2$ persistence. 
In {\bf b}, {\bf c} and {\bf d},  the horizontal axises are the density isovalues (i.e., the main filtration parameter).  The vertical axises represent the scale (\AA).   Color bars  denote PBNs. 
}
\label{fig:sigma_c60}
\end{figure}
 
We generate volumetric density data of   fullerene C$_{60}$   by using the exponential function
\begin{eqnarray}
\mu({\bf r})  = \sum_{j=1}^N  e^{-\frac{\|{\bf r} - {\bf r}_j \|}{\sigma}},
\end{eqnarray}
where the scale $\sigma$ is utilized as a multiscale parameter and will be varies from 0.2\AA~ to 1.7\AA. For each given scale, we carry out the density isovalue  based filtration of fullerene C$_{60}$.  Our results are depicted in Fig. \ref{fig:sigma_c60}. A C$_{60}$ surface is plotted in Fig. \ref{fig:sigma_c60} {\bf a}. Clearly C$_{60}$ has 60 carbon atoms which gives rise to 60 $\beta_0$  bars. Additionally, these atoms can form 12 pentagons and 20 hexagons, which lead to 31 1D rings or 31 $\beta_1$ bars. Moreover, C$_{60}$ can form a central cavity, which contributes to a $\beta_2$ bar. Finally, our earlier study shows that 20 hexagons also produces 20 Betti-2  topological invariants due to the use of  Vietoris $-–$ Rips complex \cite{KLXia:2014c}. The present work examine the topological persistence under different scales.  Figures \ref{fig:sigma_c60} {\bf b}, {\bf c} and {\bf d} illustrate respectively  2D $\beta_0$, $\beta_1$ and $\beta_2$ persistence over the scale change. The horizontal and  vertical axises represent the density and scale, respectively. Color bars denote PBNs. In our persistent homology analysis, the filtration goes from the highest density value of 7.6 to the lowest value of 0.0.  The scale varies over range [0.2\AA, 1.7\AA]. 

The impact of scale to topological invariants is very dramatic. Since isolated entities exist typically at small scale, $\beta_0$ values are relatively large at small $\sigma$ values, i.e.,  0.2\AA~ to 0.5\AA. In contrast, one dimensional rings are favored at the intermediate scale (0.4\AA ~to 1.0\AA). Therefore, as the scale is increased,   $\beta_1$ values increase first then decay at large $\sigma$ values. Since the cavity of C$_{60}$ is a relatively large scale property,  $\beta_2$ values are enhanced at relatively large scales (0.6\AA ~ to 1.2\AA). However, when $\sigma$ value goes beyond the radius of 1.4\AA, all topological invariants diminish. In summary, the scale $\sigma$ works as a control parameter  for ``the focus of  the optical len'' of persistent homology.  If one needs to identify local isolated structures, one just selects a small $\sigma$  value. If one concerns the global patterns, one increase the scale to appropriate size. However,  when the scale  goes beyond the geometric dimension, the structure gets blurred.

\paragraph{Multiscale high-dimensional  persistence}

\begin{figure}
\begin{center}
\begin{tabular}{c}
\includegraphics[width=0.8\textwidth]{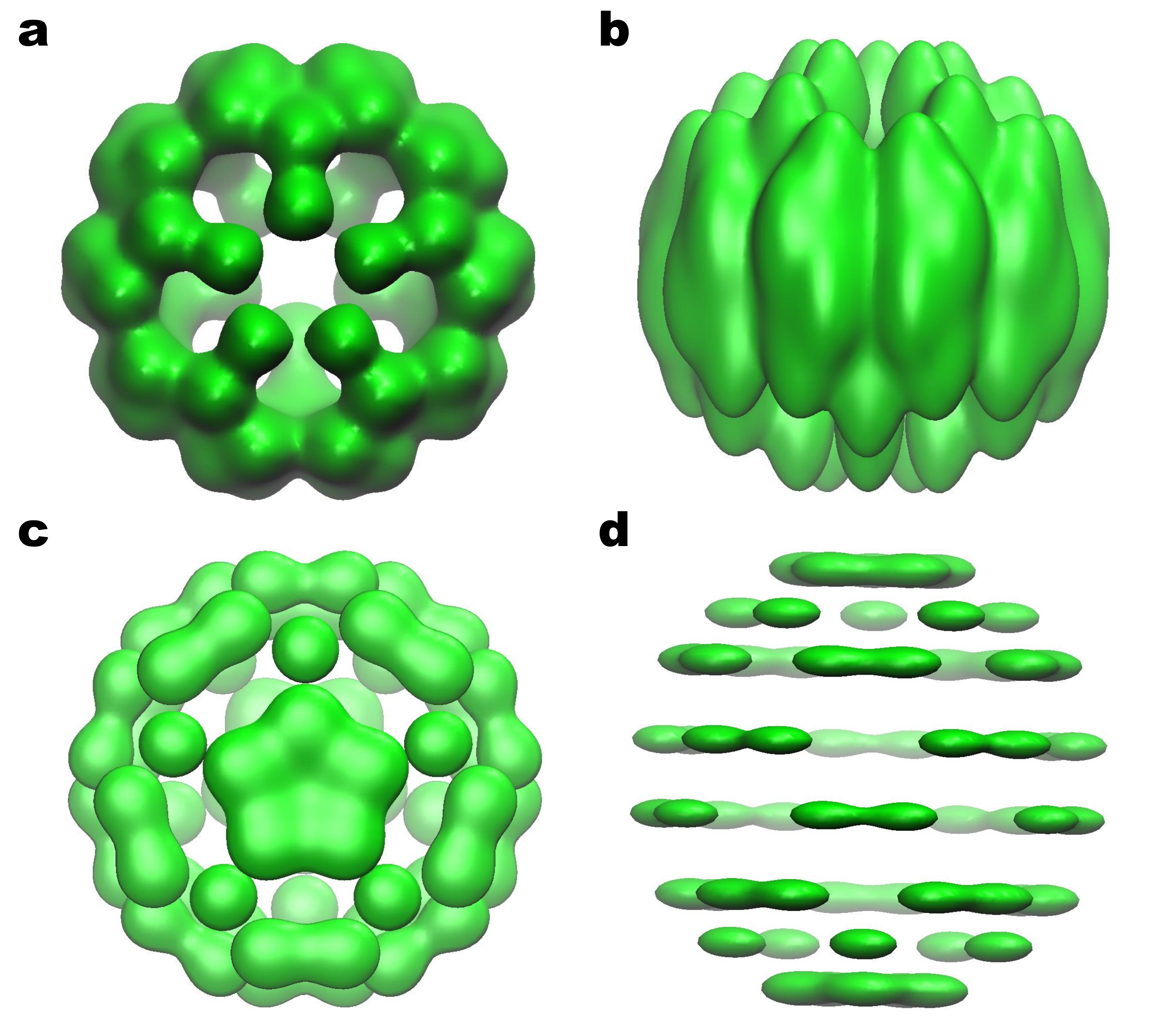}
\end{tabular}
\end{center}
\caption{ Multidimensional anisotropic filtration of C$_{60}$.  
{\bf a} The $z$-direction view of C$_{60}$ with $\sigma^x=\sigma^y=0.2$ \AA~ and $\sigma^z=0.5$ \AA ~ at the isovalue of 0.4; 
{\bf b} The $x$-direction view of C$_{60}$ with $\sigma^x=\sigma^y=0.2$ \AA~ and $\sigma^z=0.5$ \AA ~ at the isovalue of 0.4; 
{\bf c} The $z$-direction view of C$_{60}$ with $\sigma^x=\sigma^y=0.5$ \AA~ and $\sigma^z=0.2$ \AA ~ at the isovalue of 1.0; 
{\bf d} The $x$-direction view of C$_{60}$ with $\sigma^x=\sigma^y=0.5$ \AA~ and $\sigma^z=0.2$ \AA ~ at the isovalue of 1.0. 
}
\label{fig:3D_c60}
\end{figure}

Having demonstrated the construction of 2D topological persistence in a number of ways, we pursue  to the development of 3D persistence. Obviously, there are a variety of ways that one can construct 3D or multidimensional persistent homology. For example, 3D persistent homology can be generated by the combination of scale, time and  the matrix filtration, the combination of scale, time and density filtration, and the combination of scale, SNR and density filtration. In the present work, we illustrate 3D persistent homology  by using anisotropic scales or anisotropic filtrations,  which give rise to truly  multidimensional simplicial complexes and truly  multidimensional persistent homology.  For simplicity, we still take fullerene C$_{60}$ as an example to illustrate our approach.

We define the density of the fullerene C$_{60}$ by a multiscale function,
\begin{eqnarray}\label{C60density}
\mu({\bf r}) =  \sum_{j=1}^{60} \frac{1.0}{1.0+\sqrt{(\frac{x-x_j}{\sigma_j^{x}})^2+ (\frac{y-y_j}{\sigma_j^{y}})^2+ (\frac{z-z_j}{\sigma_j^z})^2 }},
\end{eqnarray}
where $(x_j,y_j,z_j)$ are the atomic coordinates of C$_{60}$ molecule and  $\sigma_j^x, \sigma_j^y$ and  $\sigma_j^z$ are 180 independent scales.  Obviously, each of these scales can vary  independently. Therefore,  together with  the density, these scales are able to deliver 181-dimensional filtrations. However, the visualization of such a high-dimensional persistent homology will be a problem, not to mention its physical meaning.  To reduce the dimensionality, we set $\sigma_j^x= \sigma^x$, $\sigma_j^y= \sigma^y$ and $\sigma_j^z= \sigma^z$, which leads to four-dimensional (4D) persistent homology. To further reduce the dimensionality, we set $\sigma^x=\sigma^y$ to end up with 3D  persistence.

\begin{figure}
\begin{center}
\begin{tabular}{c}
\includegraphics[width=0.8\textwidth]{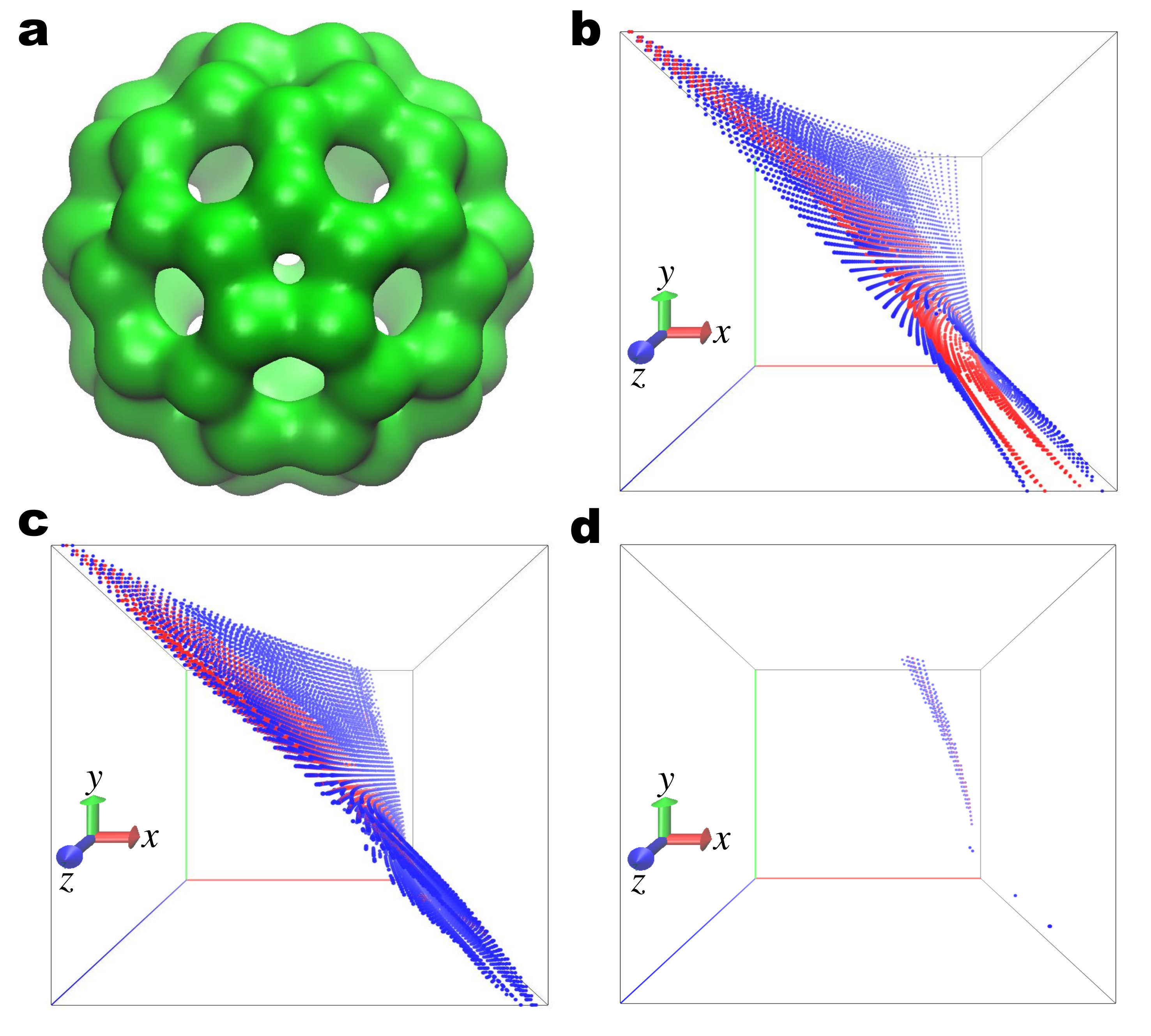}
\end{tabular}
\end{center}
\caption{The C$_{60}$ molecule and its multiscale 3D persistence. 
{\bf a}   C$_{60}$ molecule obtained with  $\sigma^{x}=\sigma^{y}= \sigma^{z}=0.5$ \AA ~ at the isovalue of 1.5;
{\bf b} 3D $\beta_0$ persistence;
{\bf c} 3D $\beta_1$ persistence;  
{\bf d} 3D $\beta_2$ persistence. 
In {\bf b}, {\bf c} and {\bf d}, the  $x$-axises label the density value (i.e., the main filtration parameter), the $y$-axises denote $\sigma^z$ and the $z$-axises represent   $\sigma^{x}=\sigma^{y}$. The blue and red dots denote $\beta_0=4$ and 50 respectively  in {\bf b},   $\beta_1=3$ and 20 respectively  in {\bf c}, and 
 $\beta_2=1$ and 2 respectively  in {\bf d}. 
}
\label{fig:3Dpersistence_c60}
\end{figure}

Unlike the isotropic filtration created by an isotropic scale,  the anisotropic filtration  creates a family of distorted ``molecules'' for topological analysis.  For the highly symmetry C$_{60}$ molecule, these distorted versions are not very physical by themselves. However, C$_{60}$ is a good choice for illustrating and analyzing our methodology, because any distortion is due to the method. On the other hand, the method itself is meaningful due to the fact that  most molecules are not symmetric and have anisotropic shapes or     anisotropic thermal fluctuations. Figure \ref{fig:3D_c60} depicts anisotropic  C$_{60}$ molecules generated by different combinations of  $\sigma^x=\sigma^y$ and  $\sigma^z$ according to Eq. (\ref{C60density}). Figures \ref{fig:3D_c60} {\bf a} and {\bf b} are obtained with $\sigma^x=\sigma^y=0.2$ \AA~ and $\sigma^z=0.5$ \AA ~ at the isovalue of 0.4. There is an elongation along the $z$ axis.  Figures \ref{fig:3D_c60} {\bf c} and {\bf d} are generated  with $\sigma^x=\sigma^y=0.5$ \AA~ and $\sigma^z=0.2$ \AA ~ at the isovalue of 1.0. In this case,  there is an obvious compression in the $z$-direction.

Topologically, the anisotropic filtration systematically creates a family of truly multidimensional simplicial complexes which would be difficult to imagine otherwise in the 3D space.   Figure \ref{fig:3Dpersistence_c60} illustrates the multiscale 3D persistent homology of C$_{60}$ molecule.  The molecular structure is presented in Fig. \ref{fig:3Dpersistence_c60} {\bf a} with  $\sigma^{x}=\sigma^{y}= \sigma^{z}=0.5$ \AA ~ at the isovalue of 1.5. For the 2D persistent homology, the variation of PBNs over two axises  can be represented by different color schemes. However,  the visualization of PBNs in 3D is not trivial.  Figures \ref{fig:3Dpersistence_c60} {\bf b},  {\bf c} and {\bf d} are respectively multiscale 3D $\beta_0$, $\beta_1$ and $\beta_2$ persistence. Here the $x$-axises represent the density value (i.e., the main filtration parameter). The $y$-axises denote $\sigma^z$ and the $z$-axises are for  $\sigma^x =\sigma^y$. The distributions of two PBNs, $\beta_0=4$ and  $\beta_0=50$ are plotted with blue dots and red dots respectively in  Fig. \ref{fig:3Dpersistence_c60} {\bf b}. It is seen that PBNs of $\beta_0$ are mainly distributed at small  $\sigma^x$ and $ \sigma^z$ scales, which is consistent with the findings obtained in the previous multiscale 2D persistence shown in    Fig. \ref{fig:sigma_c60} {\bf b}.  In  Fig.  \ref{fig:3Dpersistence_c60} {\bf c}, we depict the distributions of $\beta_1=3$ and  $\beta_1=20$ with blue dots and red dots, respectively. Our 3D results are quite similar to those given in Fig. \ref{fig:sigma_c60} {\bf c}.  As the scales increase,  the PBNs of $\beta_1$  first increase then decay.   Finally,   the distributions of $\beta_2=1$ and  $\beta_2=2$ are illustrated with blue dots and red dots, respectively in  Fig.  \ref{fig:3Dpersistence_c60} {\bf d}. As the cavity of C$_{60}$ is relatively global, the values of $\beta_2=1$ is seen to locate at relatively large scales. This result is consistent with that revealed in  Fig. \ref{fig:sigma_c60} {\bf d}.

\section{Conclusion}

Recently, persistent homology, a new branch of topology, has gained considerable popularity for computational application in big data simplification. It generates a one-parameter family of topological spaces via filtration such that topological invariants can be measured at a variety of geometric scales. As a result, persistent homology is able to bridge the gap  between geometry and topology. However, one-dimensional (1D) persistent homology has its limitation to represent high dimensional complex data. Multidimensional persistence, a generalization of 1D persistent homology to a multidimensional one,  provides a new promise for big data analysis. Nevertheless, the realization and construction of robust multidimensional persistence have been a challenge. 

In this work, we introduce two types of multidimensional persistence.  The first type  is called pseudo-multidimensional persistence, which is generated by the repeated applications of 1D persistent homology to high-dimensional data, such as results from molecular dynamics simulation, partial differential equations (PDEs), molecular surface evolution, video data sets, etc. The other type of multidimensional persistence is constructed by appropriate multifiltration processes. Specifically, cutoff distance and scale are introduced as new filtration variables to create multifiltration and multidimensional persistence. The scale of flexibility-rigidity index (FRI) \cite{KLXia:2013d,Opron:2014} behaves in the same manner as the wavelet scale.  It serves as an independent  filtration variable and controls the formation of simplicial complexes and the corresponding  topological spaces. As a result, the FRI scale creates  truly multiscale multidimensional persistent homology, in conjugation with    the matrix value variable or the density  variable. We have developed  genuine two-dimensional (2D) persistent homology. By using anisotropic scales, in which the scale in each spatial direction can vary independently, we can  construct four dimensional (4D) persistent homology. A protocol is prescribed  for the construction of arbitrarily high dimensional persistence.  Concrete numerical example is given to three-dimensional (3D) persistence.

We have demonstrated the utility, established the robustness and explored the efficiency of  the proposed  multidimensional persistence by its applications to a wide range of biomolecular systems. First, we have constructed pseudo-multidimensional persistence for the protein unfolding process. It is shown that local  topological features such as pentagonal and hexagonal rings in the amino acid residues are preserved during the unfolding process, whereas global topological invariants diminish over the unfolding process.  Topological transition from folded or partially folded proteins to unfolded proteins can be clearly identified in the 2D persistence. We show that  the $\beta_0$ persistence also provides an indication of the strength of applied pulling forces in the  steer molecular dynamics. Additionally, we have analyzed the optimal cutoff distance of the Gaussian network model (GNM) and the optimal scale of the FRI theory by using 2D persistence.  We have revealed the relationship between the topological connectivity in terms of Betti numbers  and the performance of the GNM and the FRI for the prediction of protein  Debye–Waller factors. Moreover, we have utilized 2D persistence to illustrate the topological signature of Gaussian noise. The efficiency of Laplace-Beltrami flow based topological denoising is studied by the present  2D persistence. We show that the topological invariants of C$_{20}$, especially $\beta_2$, persist during the denoising process, whereas the topological invariants of noising diminish during the denoising process. Similar results are also observed for the topological denoising of cryo-electron microscopy (cryo-EM) data.  Finally, we have employed multiscale multidimensional persistence to investigate the topological behavior of C$_{60}$ over isotropic and anisotropic scale variations. This study unveils that  $\beta_0$ invariants are intrinsically local, while  $\beta_1$ and $\beta_2$ invariants are relatively global. 

Multidimensional persistence techniques have been developed for three types of data formats, i.e., point cloud data, matrix data and volumetric data. We have also illustrated conversion of point cloud data to matrix and volumetric data via the FRI theory. Therefore, the proposed methodology can be directly applied to other biomolecular systems, biological networks, and   diverse  other disciplines.

\section*{Acknowledgments}

This work was supported in part by NSF grants  DMS-1160352 and IIS-1302285,  NIH Grant R01GM-090208 and MSU  Center for Mathematical Molecular Biosciences initiative. The authors acknowledge the Mathematical Biosciences Institute for hosting valuable workshops.

\vspace{0.6cm}


\end{document}